\documentclass{aa}
\usepackage{graphicx}

\newcommand\simlt{\lower.5ex\hbox{$\; \buildrel < \over \sim \;$}}
\newcommand\simgt{\lower.5ex\hbox{$\; \buildrel > \over \sim \;$}}

\newcommand\al{\alpha}

\newcommand\be{\begin{equation}}
\newcommand\ee{\end{equation}}
\newcommand\ba{\begin{eqnarray}}
\newcommand\ea{\end{eqnarray}}

\newcommand\rs[1]{_\mathrm{#1}}

\newcommand\Rcd{R\rs{cd}}
\newcommand\Rcdz{R\rs{cd,\it o}}
\newcommand\bRcd{\bar R\rs{cd}}
\newcommand\Rts{R\rs{ts}}
\newcommand\Msh{M\rs{sh}}
\newcommand\rhoej{\rho\rs{ej}}
\newcommand\Lz{L_o}
\newcommand\alz{{\al_o}}
\newcommand\Gm{\Gamma}

\begin{document}

\title{The effects of spin-down on the structure and evolution\\
of pulsar wind nebulae}

\author{ 
N. Bucciantini \inst{1,2}, R. Bandiera\inst{3}, J.~M. Blondin\inst{2}, E. Amato \inst{3}, L. Del Zanna \inst{1}}

\offprints{N.Bucciantini\\
 e-mail: niccolo@arcetri.astro.it}

\institute{Dip. di Astronomia e Scienza dello Spazio,
  Universit\`a di Firenze, Largo E.Fermi 2, I-50125 Firenze, Italy
\email{niccolo@arcetri.astro.it}\\
\and
Department of Physics, North Carolina State University,
Raleigh, NC 27695, USA\\
\and
INAF, Osservatorio Astrofisico di Arcetri,
 Largo E.Fermi 5, I-50125 Firenze, Italy
}
\authorrunning{N. Bucciantini et al.}
\titlerunning{The effects of spin-down on the structure and evolution\\
of PWNe}

\date{Received 25 September 2003; accepted 16 April 2004}
\abstract{We present high resolution spherically symmetric relativistic magnetohydrodynamical simulations of the evolution of a pulsar wind nebula inside the free expanding ejecta of the supernova progenitor. The evolution is followed starting from a few years after the supernova explosion and up to an age of the remnant of 1500 years. We consider different values of the pulsar wind magnetization parameter and also different braking indices for the spin-down process. We compare the numerical results with those derived through an approximate semi-analytical approach that allows us to trace the time evolution of the positions of both the pulsar wind termination shock and the contact discontinuity between the nebula and the supernova ejecta. We also discuss, whenever a comparison is possible, to what extent our numerical results agree with former self-similar models, and how these models could be adapted to take into account the temporal evolution of the system. The inferred magnetization of the pulsar wind could be an order of magnitude lower than that derived from time independent analytic models.
}

\maketitle

\section{Introduction}
\label{sec:intro}

Pulsars are rapidly spinning magnetized neutron stars that usually form as the result of the core collapse of massive stars ($8-16\ M_\odot$) in supernova events (SN). The typical energy released in a supernova explosion is of order $\sim10^{53}$ erg. Most of this is carried away by neutrinos, while only a small fraction (about 1\%) goes into a blast wave that sweeps up the outer layers of the star and produces a strong shock propagating in the surrounding medium. The ejected material is initially heated by the blast wave and set into motion. Then, while the heat is converted into kinetic energy, the ejecta accelerate until the pressure becomes so low as to be dynamically unimportant. When this happens the material finally sets into homologous expansion (\cite{chevalier89}; \cite{matzner99}). This phase is usually referred to as free expansion of the ejecta.

As a consequence of the electromagnetic torques acting on it, the pulsar supplies a late energy input to the remnant in the form of a relativistic magnetized wind, mainly made of electron-positron pairs and a toroidal magnetic field (\cite{goldreich69}; \cite{michel99}). Most of the pulsar rotational energy is carried away by this wind, whose propagation velocity is ultra-relativistic, with typical Lorentz factors that, far enough from the light cylinder, are estimated to be in the range $~10^{4}$--$~10^{7}$. The interaction of the wind with the ejecta expanding at non-relativistic speed produces a reverse shock that propagates toward the pulsar (\cite{rees74}). In the region bound by the wind termination shock on the inner side, and by the ejecta on the outer side, a bubble of relativistically hot magnetized plasma is created. This shines through synchrotron and Inverse Compton emission in a very broad range of frequencies, from radio wavelengths up to gamma rays: this is what we call a pulsar wind nebula (PWN) or plerion.

The evolution of a PWN inside the free expanding ejecta depends on many different parameters such as the pulsar luminosity, the density and velocity distribution in the SN ejecta (\cite{dwarkadas89}; \cite{featherstone01}; \cite{blondin96}), the presence of large and/or small scale inhomogeneities (\cite{chevalier89}; \cite{campbel03}). In the case of constant luminosity, and if spherical symmetry is assumed, it is possible to derive a simple evolutionary equation for the radius of the PWN as a function of time, that results in a power law (\cite{chevalier92}; \cite{swaluw01}). In the case of SN ejecta with a constant density profile the PWN contact discontinuity evolves as $t^{6/5}$. For a more detailed description of the various phases of the PWN-SNR evolution see, for example, van der Swaluw et al.\ (2001) and references therein.

While many analytic and numerical models exist in the literature for the evolution of SNRs, until recently only two classes of analytic models in the proper relativistic magnetohydrodynamical regime have been presented for PWNe: the steady state solution by \cite{kennel84} (KC hereafter), and the self-similar solution by \cite{emmering87} (EC hereafter), which allows for a non zero velocity of the termination shock.

Only lately the evolution of PWN-SNR systems has began being investigated through numerical simulations. These have been performed mainly in the classical hydrodynamical (HD) (\cite{blondin01}; \cite{swaluw01}), or classical MHD (\cite{swaluw03}) regime. However the recent development of codes for relativistic magnetohydrodynamics (RMHD) allows one to investigate such systems in a proper regime and to quantify the accuracy of approximate analytic solutions (\cite{bucciantini03}).

Both KC and EC models rely on two strong assumptions: a constant pulsar spin-down luminosity, and a constant velocity at the outer boundary of the nebula, neither of which applies to a real case nor is consistent with the PWN evolution inside free expanding ejecta. While both assumptions are known to be unrealistic, the most crucial one, as far as the long-term evolution of the system is concerned, is probably that of constant pulsar energy input. As we have already mentioned, the PWN is powered by the rotational energy lost by the star due to electromagnetic braking, and this loss translates into an increase with time of the pulsar rotation period (\cite{lyne98}).

In the case of a dipolar magnetic field the torque exerted on the star results in the following relation between the spin-down rate and the pulsar frequency $\Omega$ (e.g. \cite{michel99}):
\be
\dot \Omega \propto -\Omega^{3};
\label{eq:omegadot}
\ee
while the power supplied to the wind changes with time $t$ as:
\be
-I \Omega \dot \Omega = L(t) = \frac{L_{o}}{(1+t/\tau)^{2}},
\label{eq:ltomega}
\ee
where $I$ is the momentum of inertia of the pulsar, $\tau$ is the characteristic spin-down time, and $L_o$ is the initial pulsar luminosity. More generally, if the field is not exactly dipolar one can write the pulsar (or wind) luminosity as:
\be
 L(t) = \frac{L_{o}}{(1+t/\tau)^{n}},
\label{eq:lt}
\ee
where $n=(\beta+1)/(\beta-1)$, with $\beta$ the braking index.

Estimated values of $L_{o}$ may be up to $10^{38}$--$10^{40}$~erg/s.   
Determining the braking index from observations is extremely complicated, as it requires detailed and precise pulsar timing over long time-spans. A measure of $n$ is presently available for four pulsars only (\cite{camilo00}). Among these, one, the Vela pulsar, has $n=6$, while all the others have $2<n<3$. The value of $\tau$ can be derived from the pulsar period and spin-down rate once $n$ is known.

In the following we present 1D high resolution numerical simulations of the structure and evolution of a PWN inside free expanding SN ejecta. We extend, for this initial stage, that lasts for about 2000-3000 years (until the reverse shock propagating in the SNR collapses on the PWN), previous work on the subject by Bucciantini et al. (2003). 

The plan of the paper is as follows.
In Section~\ref{sec:theor} we present a semi-analytic model for the evolution of the PWN radius and the pulsar wind termination shock, valid for any power-law profile of the ejecta and for any value of $n$. After briefly describing, in Section~\ref{sec:numsim}, the numerical code employed and the initial conditions of our simulations, we present, in Section~\ref{sec:numres}, the results obtained for different values of the braking index, $n=0,2,3$, and different magnetizations of the wind. We discuss these results and compare them with the expectations of the analytic EC model. A formula is derived for the spin-down factorization, and the new results are applied to the case of Crab Nebula. In Section~\ref{sec:fin} we summarize our conclusions.

\section{A semi-analytic preamble}
\label{sec:theor}
In this section we shall derive some approximate relations for the evolution of a PWN interacting with the SN ejecta: these will be of use to interpret the numerical results of Section~\ref{sec:numres}. Our main goal will be to find an expression for the time evolution of the position of the termination shock ($R_{ts}(t)$) and of the contact discontinuity ($R_{cd}(t)$).

First of all, using the results found by Bucciantini et al. (2003), we write the total (magnetic plus particle) energy inside the PWN as:
\be
E_{pwn}=4\pi\;R_{cd}^{3}P_{cd}\ ,
\label{eq:etot}
\ee
with $P_{cd}$ the total pressure at the contact discontinuity. The validity of this equation is easy to prove in the two extreme cases: if we assume for the PWN a non magnetized bubble with constant thermal pressure or a magnetically dominated bubble with magnetic field $B \propto r^{-1}$. However, the simulations we present in the following (Section~\ref{sec:numres}) support the idea that this relation holds to a very good approximation (within 0.3\% error), whatever the spatial distribution of magnetic field and thermal energy in the nebula, and for all values of $n$ considered. 

From Eq.~\ref{eq:etot}, it follows that the evolution of $E_{pwn}$, including the pulsar input and the effects of adiabatic expansion, can be written in the form:
\be
\frac{d}{dt}(4\pi\;R_{cd}^{3}P_{cd})=L(t)-4\pi\;R_{cd}^{2}P_{cd} \dot R_{cd}.
\label{eq:prin2}
\ee
Integration by parts leads to:
\be
P_{cd} (t)={1 \over 4 \pi R_{cd}(t)^4} \int_0^t L(s) R_{cd}(s) {\rm d}s\ .
\label{eq:ref1}
\ee
This latter equation makes it clear that, for given ejecta properties, once $L(t)$ is assigned, both $R_{cd}(t)$ and $P_{cd}(t)$ are uniquely determined, independently of the wind magnetization. 

What does depend on the wind magnetization is the internal pressure profile: while in a HD case the pressure just behind the shock would be almost equal to $P_{cd}(t)$, in a MHD case it will be higher. Moreover the ratio of the post-shock pressure over the wind ram pressure (just upstream of the shock) will depend on the termination shock speed. Approximate pressure balance at the shock will still hold in all cases where the shock speed is low compared to the speed of light. We can then write:
\be
{L(t) \over 4 \pi R_{ts}^2(t) c } \approx P_{ts}(t)={P_{ts}(t) \over P_{cd}(t)}
{\int_0^t L(s) R_{cd}(s) {\rm d}s \over 4 \pi R_{cd}^4(t)}\ ,
\label{eq:shockbal}
\ee
where in the latter equality we made use of the expression in Eq.~\ref{eq:ref1} for $P_{cd}(t)$. It should be noticed that both $P_{cd}$ and $P_{ts}$ are here defined as including only the contribution of thermal plus magnetic pressure in the nebula, whereas the particle ram pressure is neglected. The latter is expected to be negligible at the outer boundary of the nebula, while in a truly steady-state situation it would contribute 1/3 of the total pressure just behind the shock. 

Eq.~\ref{eq:shockbal} allows one to derive the time-evolution of the termination shock radius, once $R_{cd}(t)$ is known, and a model for the behavior of the ratio $P_{ts}/P_{cd}$ as a function of time is provided. As far as the first task is concerned, a simple analytic description of how the contact discontinuity position evolves with time can be found under the thin-layer approximation. In this approximation, $R_{cd}(t)$ can be derived from the following equation:
\begin{equation}
\int_0^t{\!\!\!\ \Rcd(t')L(t')\,{\rm d}t'}=
  \Rcd^2\!\left(\!\!\Msh\ddot\Rcd+
  \dot\Msh\!\left(\!\!\dot\Rcd-\frac{\Rcd}{t}\!\!\right)\!\!\right),
\label{eq:rino1}
\end{equation}
where $\Msh$ is the mass of the shocked ejecta. Eq.~\ref{eq:rino1} is obtained by combining (e.g. \cite{reynolds84})mass and momentum conservation with the energy conservation law of Eq.~\ref{eq:prin2}.

It is possible to derive a power series approximation of Eq.~\ref{eq:rino1}, which allows an analytic description of the evolution of $\Rcd$ at any time. It is worth stressing that the series expansion we present in the following can be applied to the case of a general braking law for the plerion-feeding pulsar (Eq.~\ref{eq:lt}), as well as to the case of a general power-law density profile of the ejecta $\rhoej\propto t^{\xi-3}r^{-\xi}$. We prefer to express the latter in terms of the enclosed mass, in the following way:
\begin{equation}
M=\frac{W}{(\al-1)(\al+1)\left(\al+(3-\xi)(\al-1)\right)}
  \left(\frac{R}{t}\right)^{3-\xi}.
\label{eq:rino2}
\end{equation}
We have chosen the factor $W$ in the above equation in such a way that the following simple expansion law holds for a constant energy input by the pulsar:
\begin{equation}
\Rcdz(t)=\left(L/W\right)^{\alz-1}t^\alz=
        \bRcd(t/\tau)^\alz,
\label{eq:rino3}
\end{equation}
where $\alz=(6+\xi)/(5+\xi)$, and the latter equality defines $\bRcd$.
In the general case of a fading $L(t)$ (described by Eq.~\ref{eq:lt}), the evolution of $\Rcd$ is not exactly a power-law. While, as shown by Eq.~\ref{eq:rino3}, at times much smaller than $\tau$ the best-fit power-law expansion has an index $\al=\alz$, at very late times the evolution of $\Rcd$ is well approximated by a linear law (i.e.\ $\al=1$).

We employ a series expansion that is capable of describing the evolution of $\Rcd$ at all times, and which has a functional form that allows the time-integral in Eq.~\ref{eq:shockbal} to be performed analytically. Let us introduce the variable $s=(t/\tau)/(1+t/\tau)$: the temporal range $[0,\infty]$ maps into the range $[0,1]$ for $s$.

Let us then express $\Rcd(t)$ by the series:
\begin{equation}
\Rcd(t)=\bRcd\frac{s^\alz}{1-s}\sum_{i=0}^\infty{c_is^i},
\label{eq:rcdt}
\end{equation}
with $c_0=1$. The factor in front of the series guarantees that, for $t\ll\tau$, $\Rcd(t)$ is well approximated by $\Rcdz(t)$, while at larger times $\Rcd(t)\propto t$.

\begin{table*}
  \caption[]{The first few coefficients $c_i$ for the series expansion in Eq.~\ref{eq:rcdt}. The polynomial expression of $c_i$ as a function of the pulsar spin-down index is given for different power-law profiles of the ejecta.} 
  $$
 \begin{array}{p{0.1\linewidth}p{0.8\linewidth}l}
    \hline
    \noalign{\smallskip}
$\xi=0$ &$c_1=0.200000 -0.044898\; n $ \\
$ $     &$c_2=0.120000 -0.045438\; n +0.004434\; n^2 $\\
$ $     &$c_3=0.088000 -0.043397\; n +0.007204\; n^2 -0.000383\; n^3 $\\
$ $     &$c_4=0.070400 -0.041023\; n +0.008888\; n^2 -0.000802\; n^3 +0.000024\; n^4 $\\
    \noalign{\smallskip}
    \hline
    \noalign{\smallskip}
$\xi=1$ &$c_1=0.250000 -0.056250\; n $\\
$ $     &$c_2=0.156250 -0.059742\; n +0.005875\; n^2 $\\
$ $     &$c_3=0.117188 -0.058696\; n +0.009905\; n^2 -0.000540\; n^3 $\\
$ $     &$c_4=0.095215 -0.056620\; n +0.012563\; n^2 -0.001176\; n^3 +0.000037\; n^4 $\\
    \noalign{\smallskip}
    \hline
    \noalign{\smallskip}
$\xi=2$ &$c_1=0.333333 -0.075269\; n $\\
$ $     &$c_2=0.222222 -0.086234\; n +0.008582\; n^2 $\\
$ $     &$c_3=0.172840 -0.088634\; n +0.015326\; n^2 -0.000864\; n^3 $\\
$ $     &$c_4=0.144033 -0.088316\; n +0.020295\; n^2 -0.002000\; n^3 +0.000069\; n^4 $\\
    \noalign{\smallskip}
    \hline
    \noalign{\smallskip}
$\xi=3$ &$c_1=0.500000 -0.113636\; n $\\
$ $     &$c_2=0.375000 -0.149268\; n +0.015168\; n^2 $\\
$ $     &$c_3=0.312500 -0.166823\; n +0.030039\; n^2 -0.001785\; n^3 $\\
$ $     &$c_4=0.273438 -0.176633\; n +0.042976\; n^2 -0.004568\; n^3 +0.000177\; n^4 $\\
    \noalign{\smallskip}
    \hline
  \end{array}
  $$	
\label{tabrino}	
\end{table*}

An advantage of this functional form is that the series converges at all times, while a simple power series of $t$ has been found not to converge for $t > \tau$. The values of the various coefficients can be easily obtained in the case of constant $L$ (i.e.\ $n=0$), where $\Rcd$ follows Eq.~\ref{eq:rino3}, which implies:
$$
\sum_{i=0}^\infty{\!\!c_is^i}\!=(1-s)^{1-\alz}\!=\sum_{i=0}^\infty{\!\left(
\frac{(-1)^i\Gm(2-\alz)}{\Gm(2-\alz-i)\Gm(1+i)}\right)s^i}
$$
(we have used the binomial expansion).

In the case of a general $n$, Eq.~\ref{eq:rino1} must be directly solved in order to obtain the coefficients $c_i$. The first ones are:
\begin{eqnarray}
c_1&=&\frac{(49-9\xi)-n(11-2\xi)}{245-94\xi+9\xi^2},\\
c_2&=&\Big[-((49-9\xi)^2(-7038+4029\xi-764\xi^2+48\xi^3)) \nonumber\\
    &&+n(-6398469+5961745\xi-2216513\xi^2+411061\xi^3 \nonumber\\
    &&-38028\xi^4+1404\xi^5)+n^2(624393-577770\xi  \nonumber\\
    &&+213273\xi^2-39259\xi^3+3604\xi^4-132\xi^5)\Big] \nonumber\\
    &&\Big/\Big[(2(49-9\xi)^2(5-\xi)^2(1173-476\xi+48\xi^2)\Big],
\label{eq:rino6}
\end{eqnarray}
while the further ones are too complicated to be listed here. However, it can be seen that the coefficient $c_i$ is a polynomial of $i$-th degree in $n$, and in Table~\ref{tabrino}, for a few choices of $\xi$, we list the values of these coefficients up to the $4^{\rm th}$ order.

We have tested these approximate analytic solutions with a numerical model, in thin-layer approximation, for the case of flat ejecta ($\xi=0$), which is the one relevant for the situations considered here. The discrepancy increases with time up to an asymptotic value of 18\%, 13\%, 11\% and 9\%, respectively, for the first to fourth degree approximations. In the range of times shorter than $3\tau$ (as in our simulations, see next section), the errors are not larger than 4.7\%, 2.2\%, 1.2\%, and 0.7\%, for approximations of increasing degree.

The expression chosen for $\Rcd(t)$ (Eq.~\ref{eq:rcdt}) contains only terms proportional to $(t/\tau)^\mu(1+t/\tau)^\nu$, and therefore the integral $\int LR\,{\rm d}t$ can be evaluated as a series of hypergeometric functions:
\begin{eqnarray}
&&\int_0^t L(t') R(t') \,{\rm d}t'=\bRcd\Lz\tau\sum_{i=0}^\infty\frac{c_i}{1+\alz+i}
\left(\frac{t}{\tau}\right)^{1+\alz+i}			\nonumber\\
&&_2F_1 \left(n-1+\alz+i,1+\alz+i,2+\alz+i;-t/\tau \right),
\label{eq:intlr}
\end{eqnarray}
where $_2F_1(a,b,c;z)$ is the hypergeometric function.

Given Eq.~\ref{eq:rcdt}, Eq.~\ref{eq:intlr} and the coefficients $c_i$, what is left to find, in order to obtain the time evolution of $\Rts$ from Eq.~\ref{eq:shockbal}, is an expression for the ratio $P_{ts}/P_{cd}$ as a function of time. This can be easily accomplished under the approximations that the PWN always adjusts itself to a quasi-steady solution and that the fluid motion is non-relativistic everywhere behind the termination shock. Let us define the quantity $\Pi \equiv P / (\rho_u c^2 \gamma^2)$, where $\rho_u$ is the matter density and $\gamma$ the Lorentz factor of the realtivistic wind at the termination shock. Following KC, one finds for $\Pi$ the approximate expression:
\be
\Pi(y)={27 \over G(y)^4}\left[2+{3 y^2 \over G(y)^2}\right],
\label{eq:press}
\ee
with $y$ a non-dimensional coordinate related to the distance from the shock $r$, and to the magnetization of the wind $\sigma$ (supposed to be low),
\be
y(r)=\sqrt{81\ \sigma \over 2}  {r \over R_{ts}} \ ,
\label{eq:yr}
\ee
and the function $G(y)$ (see KC) given by:
\begin{eqnarray}
G(y)&=&1+\left[1+y^2+\sqrt{(1+y^2)^2-1}\right]^{-1/3}+\nonumber \\
&+&\left[1+y^2+\sqrt{(1+y^2)^2-1}\right]^{1/3} \ .
\label{eq:gy}
\end{eqnarray}
We use for the magnetization parameter $\sigma$ the definition $\sigma\equiv B^2/(4 \pi \rho c^2 \gamma^2)$ with $B$ and $\rho$ the wind magnetic field and matter proper density respectively.

We emphasize that the normalized pressure profile in Eq.~\ref{eq:press} is straightforward to obtain as a solution of the steady-state MHD equations under the assumption that the post-shock bulk Lorentz factor of the fluid is $\gamma=1$ and that $\sigma \ll 1$. The first term in Eq.~\ref{eq:press} is the thermal pressure, while the second term is the magnetic contribution, as can be easily checked evaluating the expression in the proper limits of $\sigma$ and hence $y$.      

The ratio $P_{ts}/P_{cd}$ in Eq.~\ref{eq:shockbal}, which we shall indicate hereafter as $1/K$, can be expressed in terms of $\Pi$ as:   
\be
{1 \over K(t)}={P_{ts}(t) \over P_{cd}(t)}={\Pi_0 \over \Pi(y(R_{cd}(t)))}
\label{eq:kt}
\ee
with $\Pi_0=\Pi(y(R_{ts}(t))=\Pi(\sqrt{81 \sigma/2})$. It is apparent that under these simplified assumptions the ratio between the value of the pressure at the termination shock and that at the contact discontinuity depends on time only implicitly, through the ratio between $R_{ts}$ and $R_{cd}$. 

In light of Eq.~\ref{eq:kt} we can rewrite Eq.~\ref{eq:shockbal} as an implicit equation for $y_{cd}(t) \equiv y(R_{cd}(t))=\sqrt{81 \sigma/2}\ R_{cd}(t)/R_{ts}(t)$:
\be
y_{cd}^2(t)\ \Pi(y_{cd}(t))\approx {81 \sigma \over 2} {\tau c \over \bRcd}\ \Pi_0\ Q(t)\ ,
\label{eq:yt}
\ee
where $Q(t)$ can be determined using Eq.~\ref{eq:rcdt} and Eq.~\ref{eq:intlr}:
\begin{eqnarray}
Q(t)&=&\left[\sum_{i=0}^\infty{{c_i\ (t/\tau)^i \over (1+\alz+i)} } \right.\\ 
&& \left. _2F_1\left(n-1+\alz+i,1+\alz+i,2+\alz+i;-{t \over \tau} \right) \right]\nonumber \\
&&\left(t \over \tau \right)^{1-\alz} \left(1+{t \over \tau} \right)^{n-2(1-\alpha)} \left(\sum_{i=0}^\infty{c_i (t/\tau)^i \over (1+t/\tau)^i} \right)^{-2}\nonumber 
\label{eq:qt}
\end{eqnarray}
The right-hand side of Eq.~\ref{eq:yt} is then a known function of time alone. Hence the equation can be easily solved numerically to obtain the shock position as a function of time. As we shall see in the following, this approach, despite being extremely simplified, allows one to trace the evolution of the termination shock radius with an accuracy of order 15\%, once the value of $R_{cd}$ and $R_{ts}$ are known at a reference time. In fact, as already mentioned, the first equality in Eq.~\ref{eq:shockbal} holds only within a 15-25\%, and this means that Eq.~\ref{eq:yt} must be normalized to observations (or to simulations) to provide a good estimate of the position of the termination shock at any given time.

\section{Numerical simulations}
\label{sec:numsim}
The simulations we present have been performed by using the newly developed scheme by Del Zanna et al. (\cite{delzanna02}; \cite{delzanna03}). We refer the reader to the cited papers for a detailed description of the code, and of the equations and algorithms employed. This is a high resolution conservative (shock-capturing) code for 3D-RMHD based on third order accurate ENO-type reconstruction algorithms. The approximate Riemann solver employed is the two-speed HLL flux formula, which does not make use of time-consuming characteristics waves decomposition. Given the presence of very strong shocks we have used second order reconstruction, to reduce post-shock oscillations.

We have used a single fluid model, assuming an adiabatic coefficient equal to $4/3$ also for the SN ejecta. This makes the ejecta more compressible but does not change the temporal evolution of the PWN. We have chosen to use a single fluid because, in numerical simulations, the use of two different adiabatic coefficients on a contact discontinuity with a very large density jump (density may change by factors of order $10^{6}-10^{7}$), leads to the formation of spurious disturbances that tend to propagate back into the PWN (\cite{shyue98}; \cite{karni94}; \cite{kun98}; \cite{bucciantini03}). 

When a magnetic field is present the speed of a shocked wind cannot drop to zero but tends to a finite value $V_{asy}$ (see e.g. KC). If the contact discontinuity velocity is close to this asymptotic value, even small fluctuations, originating at the PWN-SNR interface, can produce substantial variations in the internal structure. 

Another source of problems is numerical diffusion at the contact discontinuity itself. This has the effect of spreading the density jump at the contact discontinuity over a few computational cells. A criterion needs then to be established for the identification of the position of the contact discontinuity ($R_{cd}$), in order to compare the simulation results with the existing analytic models. We found that a convenient choice is to identify $R_{cd}$ with the position where the fluid velocity is equal to the contact discontinuity velocity: $v(R_{cd})=V_{cd}$. This offsets $R_{cd}$ by 3--4~\% ($R_{cd}$ is 3--4 simulation cells further out) with respect to the radius at which the density jump begins.

\subsection{Initial conditions}
\label{sec:numsim1}
Simulations have been performed on a logarithmic radial grid, with 200 cells per decade. This allows one to resolve with sufficient accuracy the inner region so that the termination shock always remains inside the computational domain and injection ambiguities such as those discussed by Bucciantini et al. (2003) are avoided. At the same time, with our choice of the grid we are able to follow the system evolution from very early times (a few years) after the SN explosion and up to an advanced age, maintaining diffusion effects homogeneous throughout the evolution. We set continuous conditions at the outer boundary (zeroth order extrapolation). No radiation cooling is included.

For the free expanding ejecta we have chosen the following profiles (\cite{chevalier92}):
\ba
\rhoej=At^{-3}    ~~~~~~~~~ ~~~~~~~~~~~~~~~~~~~~\nonumber\\
v=r/t=V_{o}r/R_o, ~~~~~~~~~ R_o(t)=V_o t,
\label{eq:eqarr} 
\ea
with $A=8.7\times 10^{6}$ g s$^{3}$/cm$^{3}$ and $V_{o}=5.27\times 10^{8}$ cm/s, corresponding to an energy release in the SN explosion $E=10^{51}$ erg and ejecta having mass $M\simeq 4 M_\odot$. 
The pulsar wind is created injecting mass, momentum-energy, and a purely toroidal magnetic field in the first computational cell, with a total luminosity that depends on time as described by Eq.~\ref{eq:lt}, with $L_{o}=5 \times 10^{39}$ erg/s, and $\tau=500$ years. Three different values for the spin-down index have been used: $n=0, 2, 3$. No magnetic field is initially present in the SNR nor in the ISM.

The simulations are initialized with a 5 years old PWN surrounded by a thin shell of swept up ejecta at a radius $R_{cd}=0.022$ ly. This shell contains the ejecta material removed from the origin by the relativistic bubble. The evolution of the PWN is followed up to an age $t=3\ \tau= 1500$ years. The internal profile of the fluid in the PWN is taken from the EC solution with proper magnetization, fitted to the contact discontinuity velocity. After a short transient phase of about 10--15 years, partly due to numerical effects, the nebula relaxes to a stable configuration. Our choice of initial conditions is such that the system is not far from the self-similar solution. This allows us to avoid the long term transient that is otherwise observed when the pulsar wind is not switched on at the same time when the SN goes off (\cite{chevalier92}; \cite{jun98}). Self-similar solutions only exist when $\sigma$ is low enough so that the asymptotic velocity of the shocked wind is smaller than the contact discontinuity initial speed, $V_{asy} < V_{cd}$. 

Three wind cases have been simulated:
\begin{itemize}
\item Purely hydrodynamic wind ($\sigma=0$); 
\item Weakly magnetized wind ($\sigma = 0.0016$);
\item ``Highly'' magnetized wind ($\sigma = 0.003$), with highly meaning close to the maximum compatible with the existence of a self-similar solution.
\end{itemize}
In all cases the wind has a Lorentz factor $\gamma=100$, and $p/\rho c^2=0.01$.

\section{Numerical results}
\label{sec:numres}
\subsection{Constant luminosity case}
\label{sec:numres1}

The first step we take is the comparison of our numerical results with the analytic model by EC in the case with $n=0$. The EC model is completely determined once the wind quantities (density, pressure, magnetic field and Lorentz factor) and the termination shock velocity are known. Given the value of the termination shock speed, $V_{ts}$, the shock jump conditions are evaluated at the termination shock radius, $R_{ts}$, and then the self-similar, spherically symmetric, RMHD equations are integrated using the post-shock values as initial conditions. A singularity appears in the solution at some distance from the termination shock: the position of this singularity corresponds to the outer boundary of the nebula. 

The self-similar solution of EC requires $V_{ts}$ to be constant and equal to a given fraction of the contact discontinuity velocity $V_{cd}$, also constant:
\be
V_{ts}=\frac{R_{ts}}{R_{cd}}V_{cd}\ .
\label{eq:presc}
\ee

We notice that the latter equation can be derived directly from the general expression in Eq.~\ref{eq:shockbal}. Let us assume a constant pulsar energy input, $L=L_0$ and the evolution of $R_{cd}$ described by a power-law: $R_{cd} \propto t^\alpha$. The time-derivative of Eq.~\ref{eq:shockbal} gives: 
\be
V_{ts}= {R_{ts} \over 2 R_{cd}} \left[ 3 V_{cd} - {R_{cd} \over t} +{R_{cd}\dot K \over K} \right]\ ,
\label{eq:vts}
\ee
which, in the self-similar case (i.e. $\alpha=1$ and $dK/dt=0$), reduces to Eq.~\ref{eq:presc}.
More generally, for $R_{cd}(t)$ still described by a power-law but with an index $\alpha\neq 1$, if the variation of $K$ in time can be neglected, we find $R_{ts}\propto t^{(3\alpha-1)/2}$. While in a HD case, one will have $K \approx {\rm const} \approx 1 $ (see also Eq.~\ref{eq:press} in the limit $y \rightarrow 0$), the time-variation of $K$ will become more and more important the larger the magnetization of the wind.

\begin{figure}
\resizebox{\hsize}{!}{\includegraphics{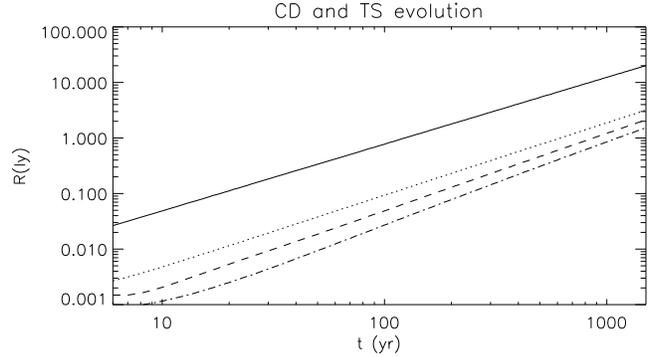}}
\caption{Evolution with time for the $n=0$ case of the position of the contact discontinuity (solid line) and of the termination shock for the hydrodynamics (dotted line), the $\sigma=0.0016$ (dashed line) and the $\sigma=0.003$ (dash-dotted line) cases.}
\label{fig:evoln0}
\end{figure}

In Fig.~\ref{fig:evoln0} we show the evolution of $R_{cd}$ derived from our simulations: this turns out to be independent of the magnetization, as stressed by Bucciantini et al. (2003). The temporal evolution agrees with the behavior $R_{cd} \propto t^{6/5}$ predicted by the analytic models (\cite{chevalier92}; \cite{swaluw01}). The value of $V_{ts}$ changes with the magnetization and is lower for larger values of $\sigma$.

We find that the evolution of the termination shock is well described by a power law in time $R_{ts} \propto t^\delta$, but the exponent changes with the wind magnetization: in the HD case we find $\delta=13/10$, as predicted by Eq.~\ref{eq:vts} for $\dot K=0$ and $\alpha=6/5$, while the value of $\delta$ increases to 1.38 and 1.43 in the cases with $\sigma=0.0016$ and $\sigma=0.003$ respectively. The evolution of the termination shock radius is very well described in these cases by Eq.~\ref{eq:yt} for the proper value of the magnetization. 
\begin{figure}
\resizebox{\hsize}{!}{\includegraphics{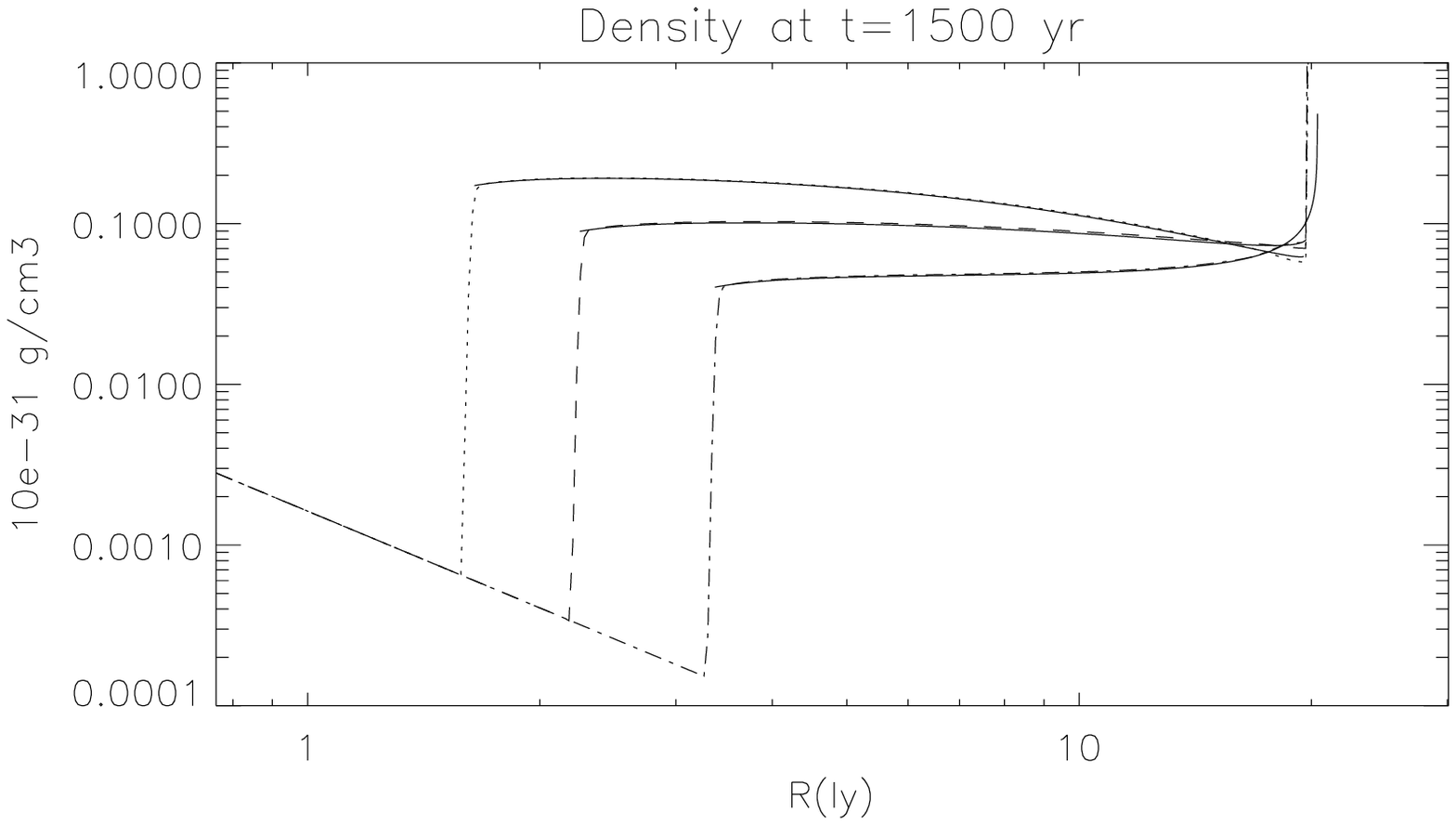}}
\resizebox{\hsize}{!}{\includegraphics{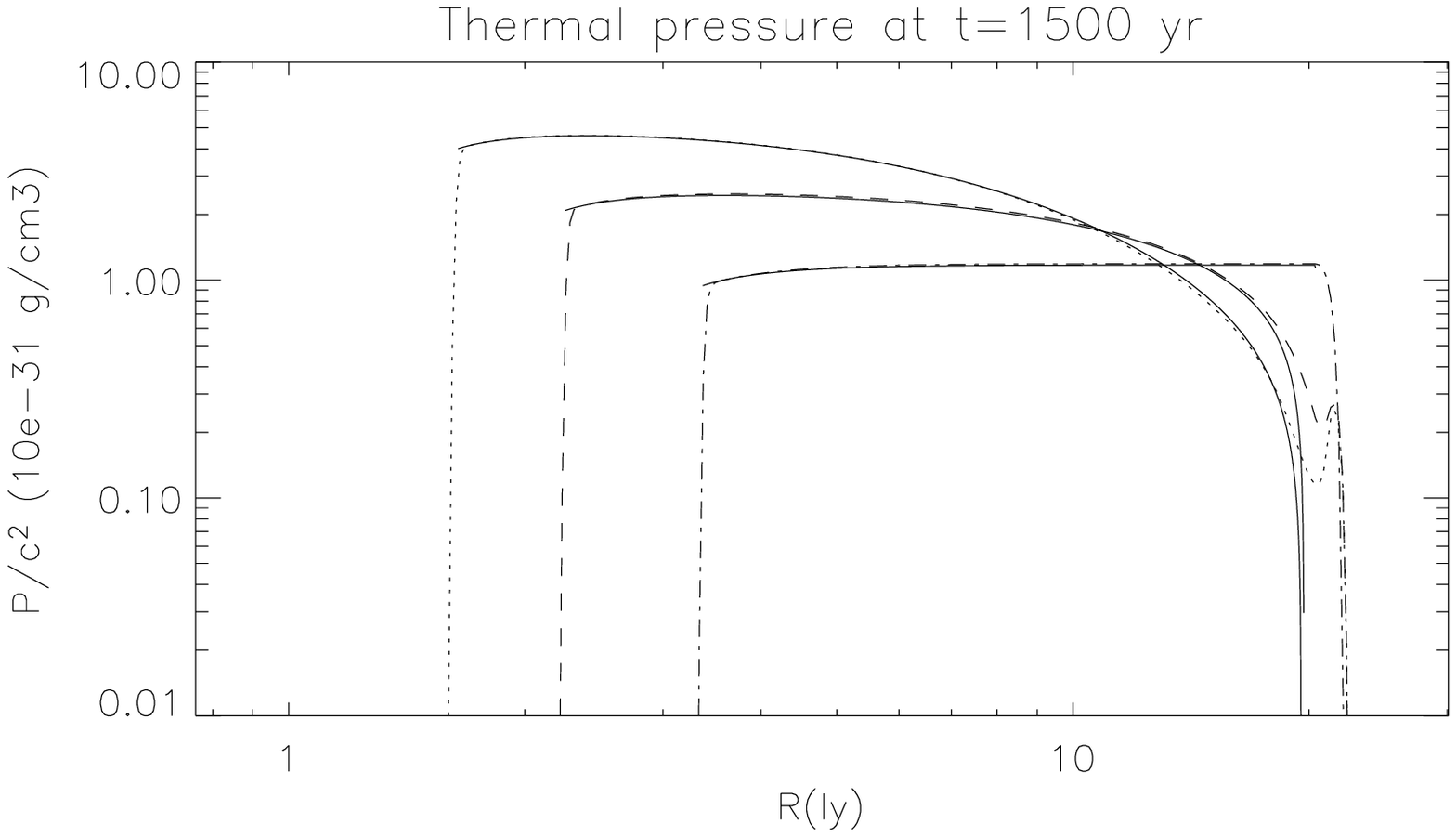}}
\resizebox{\hsize}{!}{\includegraphics{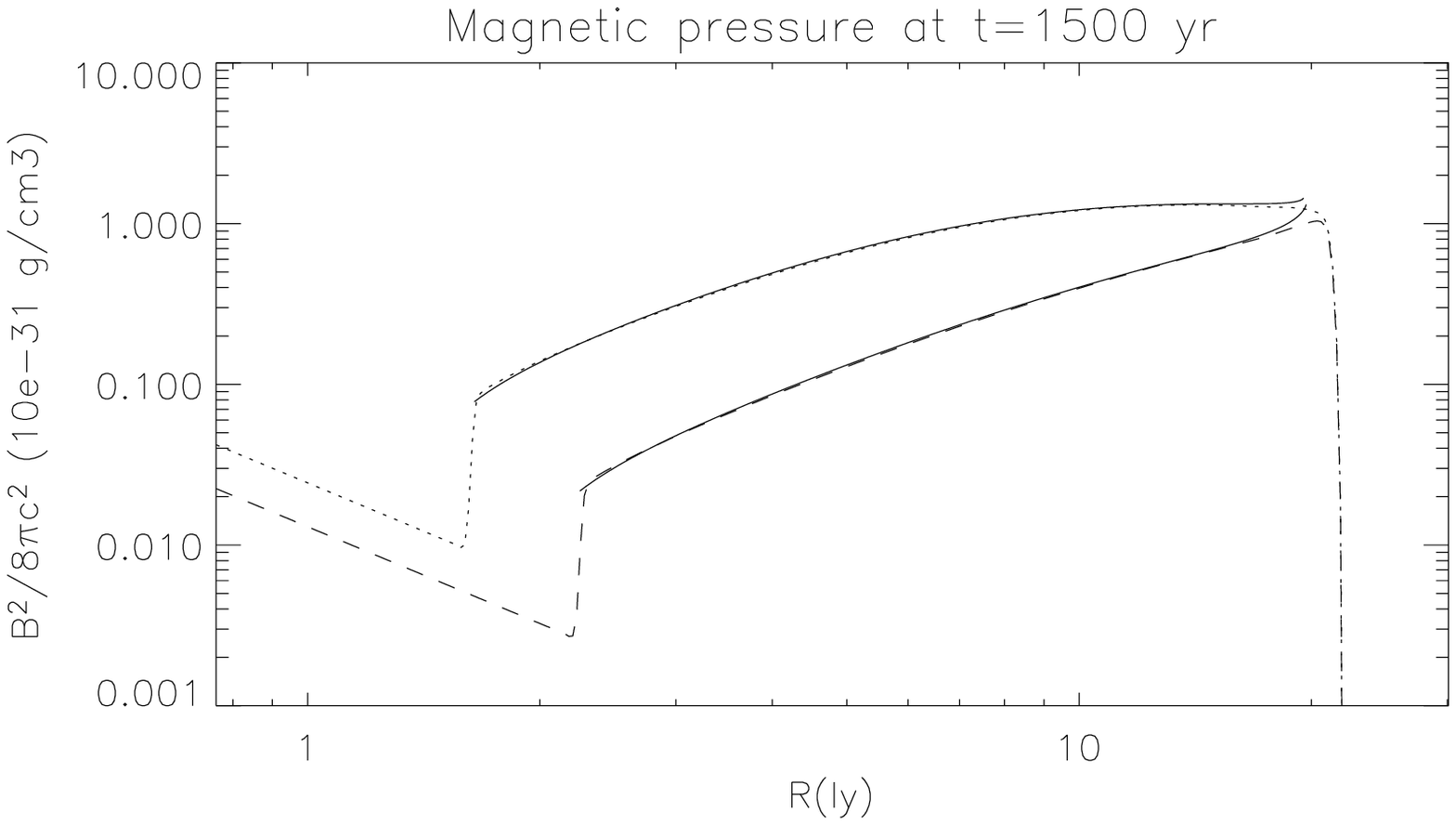}}
\caption{Comparison of the results with EC (solid line) for the no spin-down case at t=1500 yr. Density, thermal pressure and magnetic pressure are shown for all the magnetization we have considered: $\sigma=0.003$ (dotted line), $\sigma=0.0016$ (dashed line), $\sigma=0$ (dash-dotted line). The total pressure at the border is the same in the various cases. We want to stress the the radius of the contact discontinuity is about 3-4\% greater that the position where the density jump starts. This explain differences with respect to Fig.~\ref{fig:size}.}
\label{fig:profn0}
\end{figure}

We have verified that, when computing the appropriate EC model for comparison, using the value of $V_{ts}$ derived from the simulations, or that obtained from the solution of Eq.~\ref{eq:yt}, does not improve substantially the result with respect to using Eq.~\ref{eq:presc} for given values of $R_{cd}$, $V_{cd}$ and $R_{ts}$. A general advantage that Eq.~\ref{eq:presc} offers is that it allows one to estimate $V_{ts}$ from quantities that can all be measured directly, at least in principle.

In Fig.~\ref{fig:profn0} we compare the radial profiles of density and pressure (both thermal and magnetic) derived from the simulations with those computed based on EC. The agreement of the EC profiles with our results is extremely good and fails (see the pressure profile) only near the outer boundary of the nebula. This is a consequence of the self-similarity imposed in the EC model which leads to an unphysical singularity at the border. 

We want to point out some differences between the EC and KC models which lead us to conclude that the EC model is better suited as the basis for a comparison with the numerical results presented here. First of all, the EC model reproduces the positive pressure gradient in the post-shock region observed in the simulations (the pressure initially increases up to a value that is about 20-25~\% larger than that immediately behind the shock), while in KC the pressure is always monotonically decreasing. Moreover EC gives a larger flow speed in the asymptotic region than KC, in better agreement with the simulation results. 
\begin{figure}
\resizebox{\hsize}{!}{\includegraphics{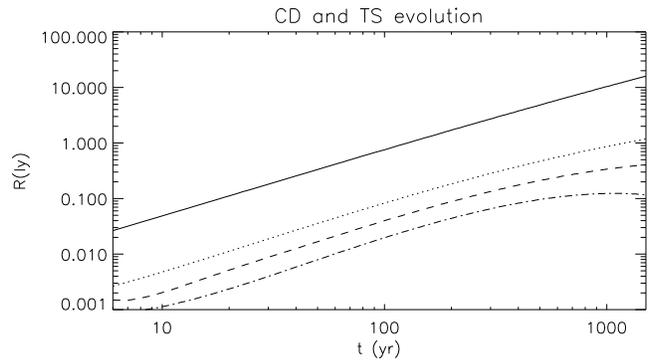}}
\caption{Evolution with time for the $n=3$ case of the position of the contact discontinuity (solid line) and of the termination shock for the hydrodynamics (dotted line), the $\sigma=0.0016$ (dashed line) and the $\sigma=0.003$ (dash-dotted line) cases. Now the evolution of the termination shock cannot be described as a power law in time.}
\label{fig:evoln3}
\end{figure}

However, the comparison with the EC model is less satisfactory if we consider the size of the nebula. In comparing the model with observations, the standard way to estimate the magnetization parameter is as follows: one finds the appropriate $\sigma$ so that the theoretical fluid speed at a distance from the shock corresponding to the observed value of $R_{cd}$ matches the boundary speed derived from observations. However, the values $R(v=V_{cd})/R_{ts}$ for the EC model are lower than the values $R_{cd}/R_{ts}$ in our simulations, and the discrepancy increases with $\sigma$ (Fig.~\ref{fig:size}). 

This discrepancy is most likely a consequence of the changing boundary speed and is enhanced when the contact discontinuity moves with a velocity close to $V_{asy}$. This is why in the HD case, when $V_{asy}=0$ (and hence $V_{cd}/V_{asy}\gg 1$) the difference is very small. On the other hand, we see in Fig.~\ref{fig:size} that, if instead of matching the velocities, we consider the size of the nebula in the EC model as the radius at which the EC solution has a singularity, we observe a much better agreement. Still some discrepancy remains in the magnetic cases but now it is well below 10~\%.

\subsection{Cases with spin-down}
\label{sec:numres2}
Let us now turn our attention to the main assumption underlying the EC solution, i.e. that of a constant pulsar luminosity. As previously noted (\cite{bucciantini03}), the spin-down process has the effect of reducing the ram pressure in the wind and, as a consequence, we expect to find a ratio $R_{cd}/R_{ts}$ greater than in the case with no spin-down.

When the effects of the pulsar spin-down are included, the evolution of $R_{cd} (t)$ can no longer be described in terms of a fixed power law in time. Despite this, we find that the variation of the exponent is small enough that we can continue to approximate it as a constant whose value, as derived from the simulations, is found to be slightly less ($\sim 1.13-1.1$) than in the case $n=0$. A correct description of the time-evolution of $R_{cd}$ can still be derived as discussed in Section~\ref{sec:theor}. Eq.~\ref{eq:rcdt} with the appropriate values of the coefficients $c_i$, as reported in Table \ref{tabrino}, is found to provide a very good approximation for $R_{cd}(t)$ within the uncertainties discussed in the same session. The same is true for $R_{ts}(t)$: this can be computed as described in Section~\ref{sec:theor} with an accuracy of order 10-15~\%, with the error increasing with increasing $n$ and $\sigma$. The reason for this can be easily understood: our approach for the computation of $R_{ts}(t)$ is based on the assumption that the evolution of the nebula proceeds slowly. This becomes an increasingly bad approximation as $n$ and $\sigma$ increase.   
\begin{figure}
\resizebox{\hsize}{!}{\includegraphics{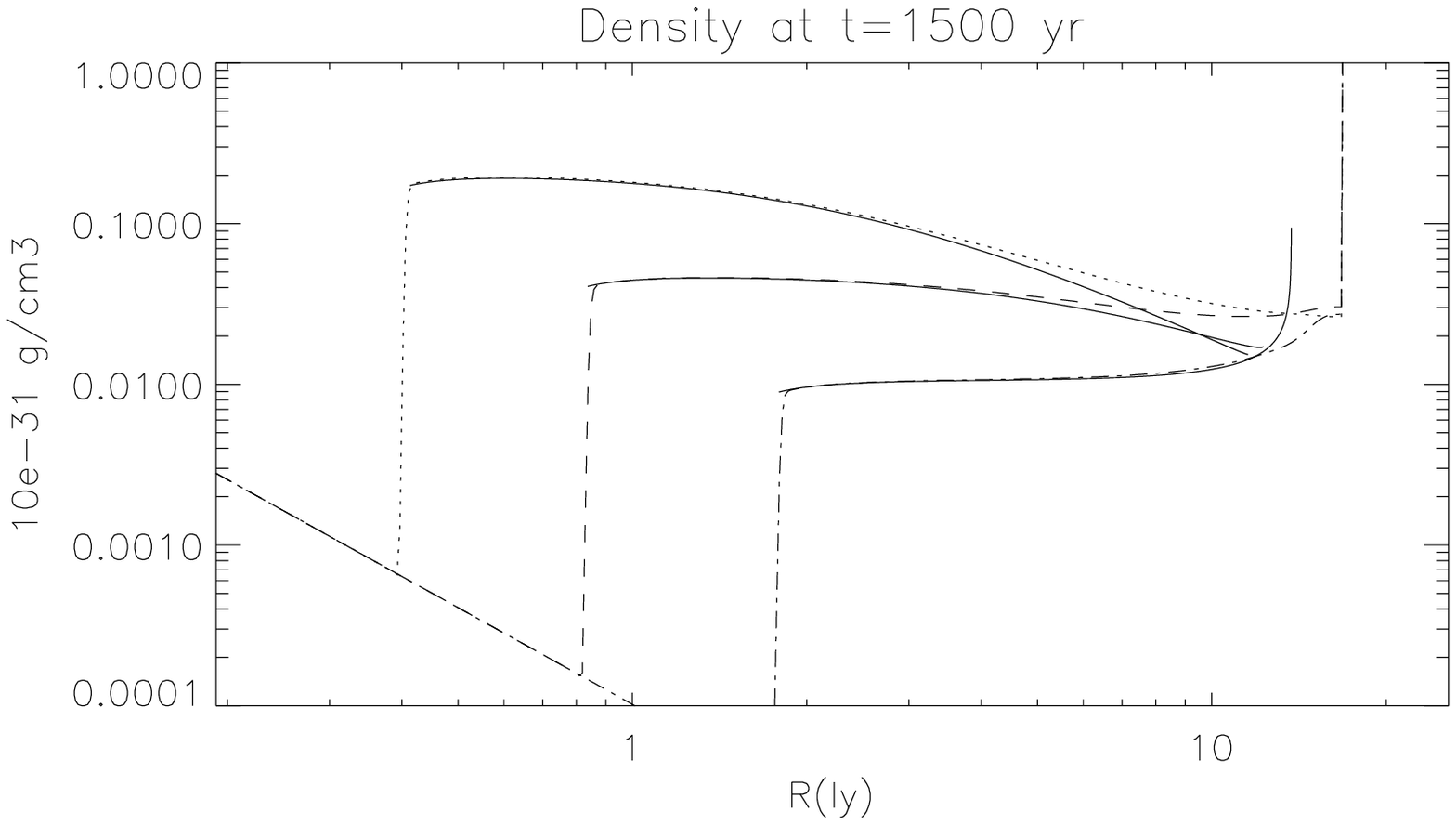}}
\resizebox{\hsize}{!}{\includegraphics{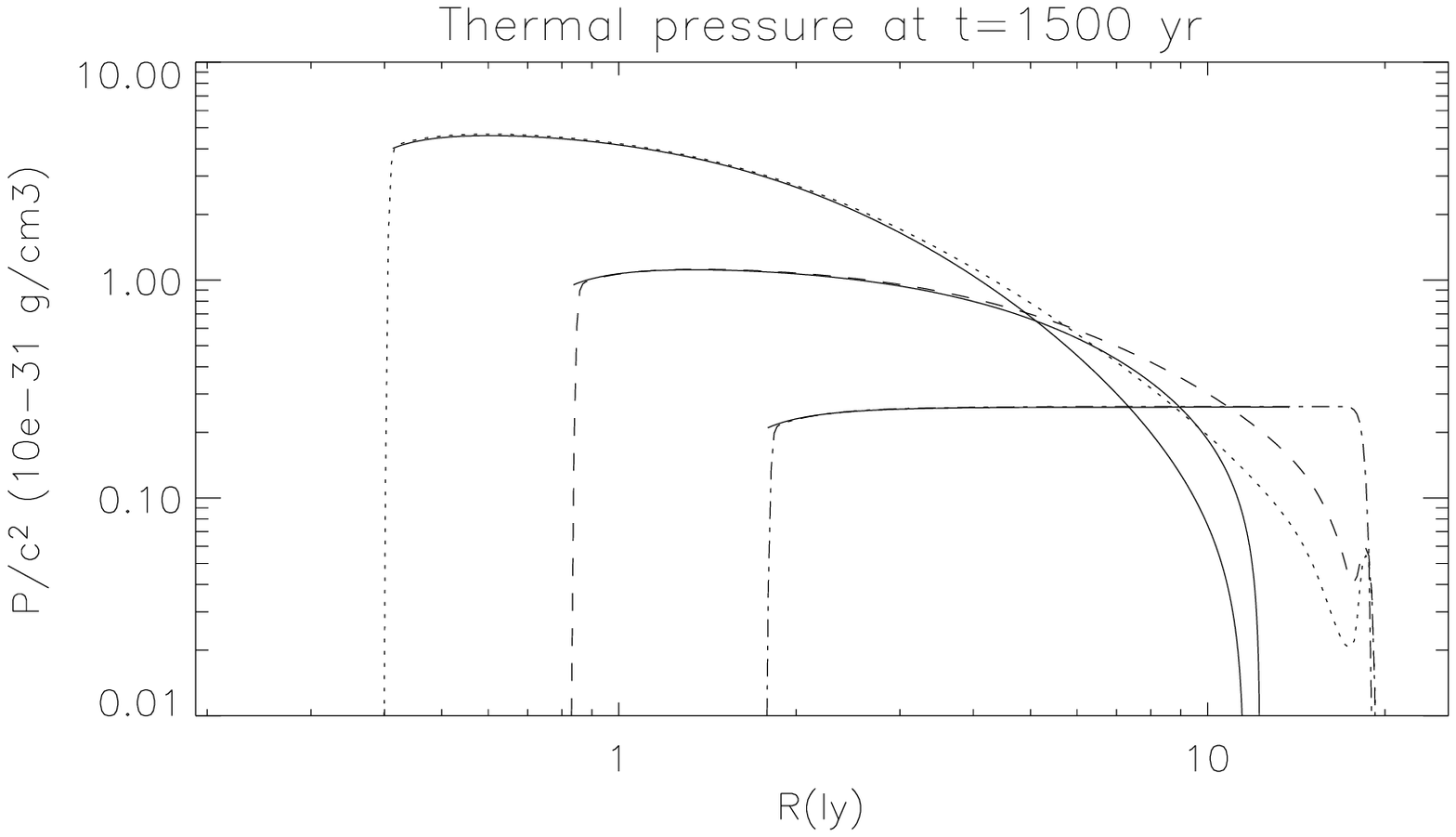}}
\resizebox{\hsize}{!}{\includegraphics{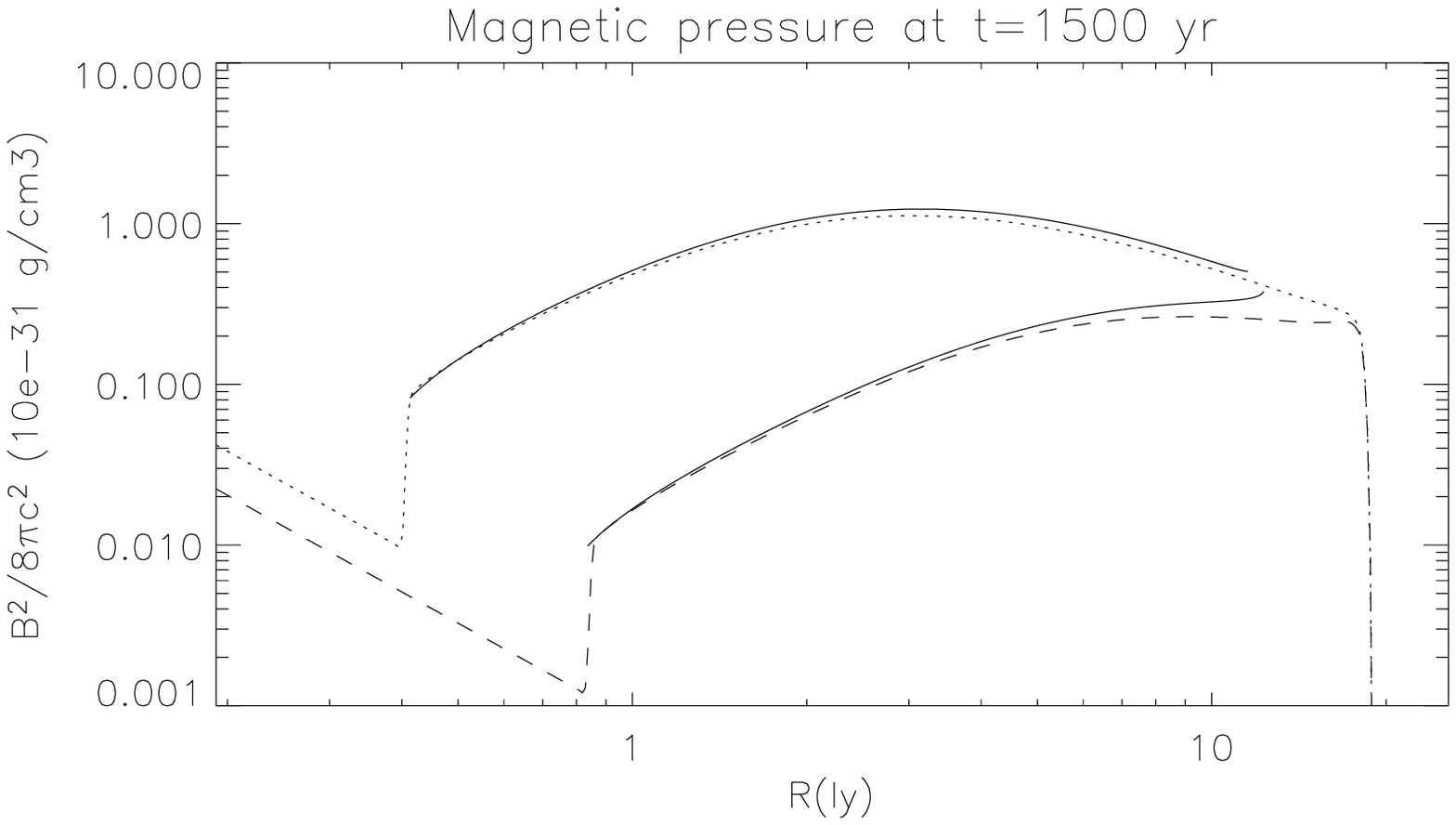}}
\caption{Comparison of the results with EC (solid line) for the $n=2$ case at t=1500 yr. Density , thermal pressure and magnetic pressure are shown for all the magnetizations we have considered: $\sigma=0.003$ (dotted line), $\sigma=0.0016$ (dashed line), $\sigma=0$ (dash-dotted line). The total pressure at the border is the same in the various cases. Comments are the same as for Fig.~\ref{fig:profn0}. }
\label{fig:profn2}
\end{figure}

\begin{figure}
\resizebox{\hsize}{!}{\includegraphics{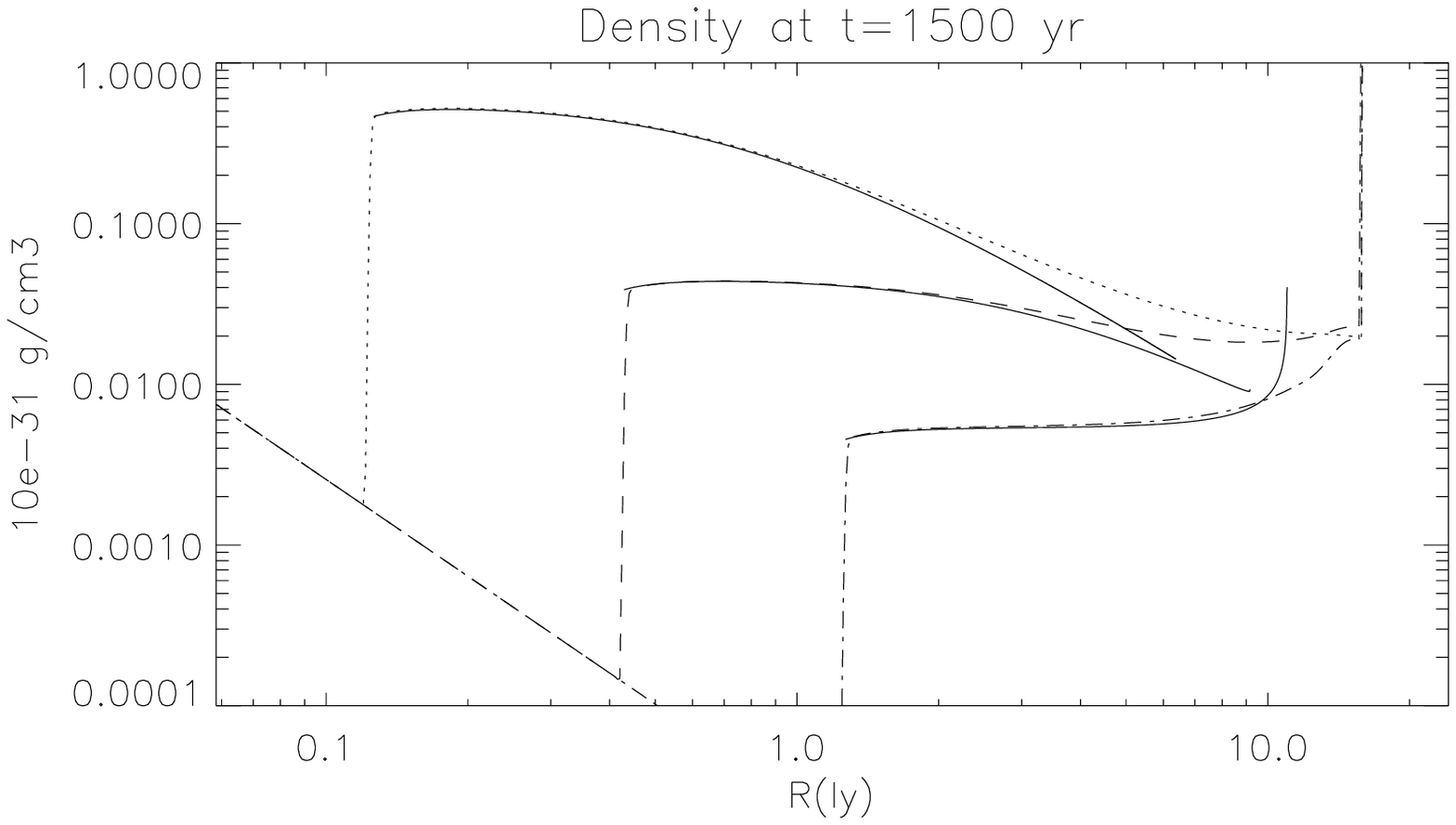}}
\resizebox{\hsize}{!}{\includegraphics{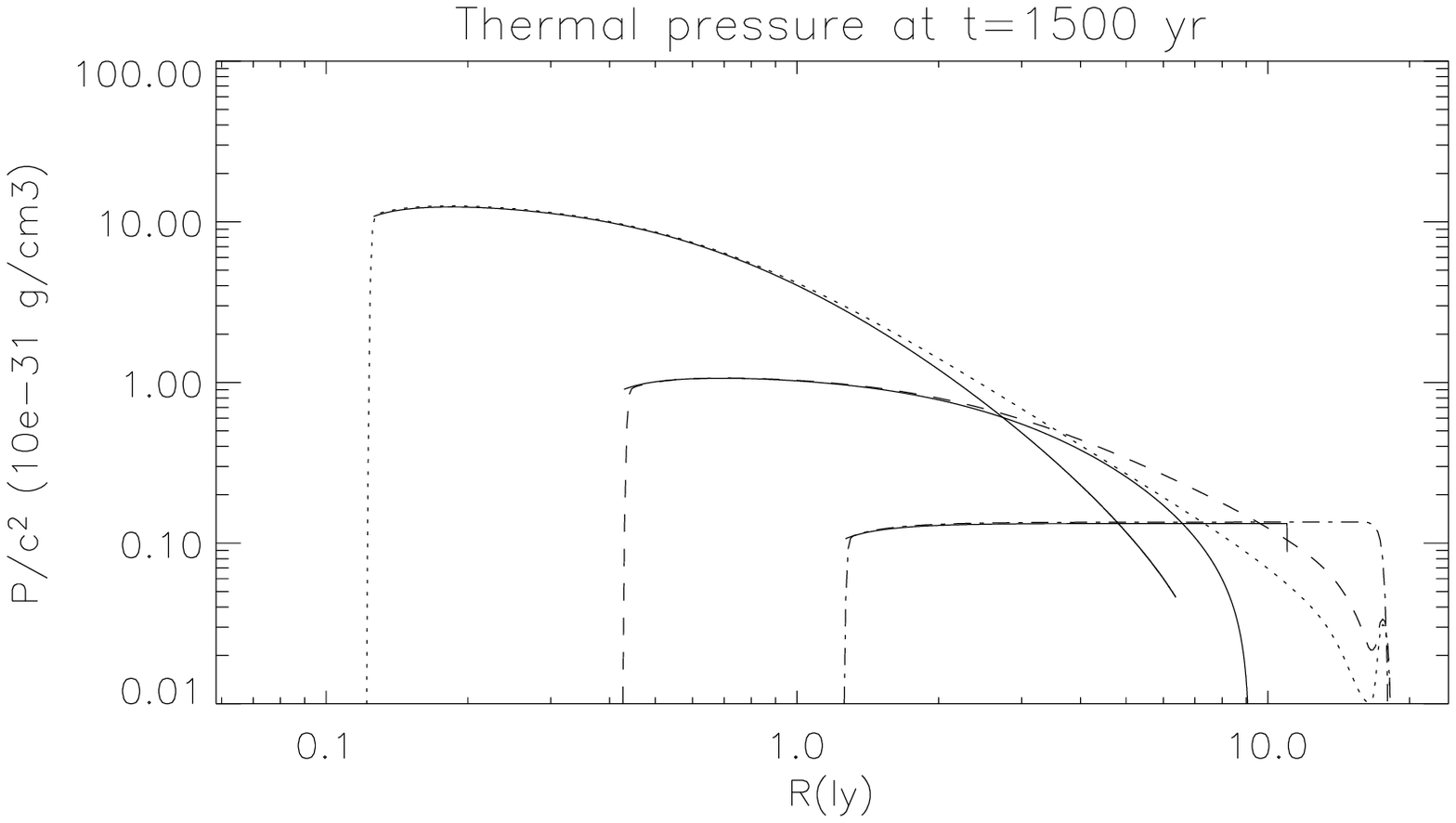}}
\resizebox{\hsize}{!}{\includegraphics{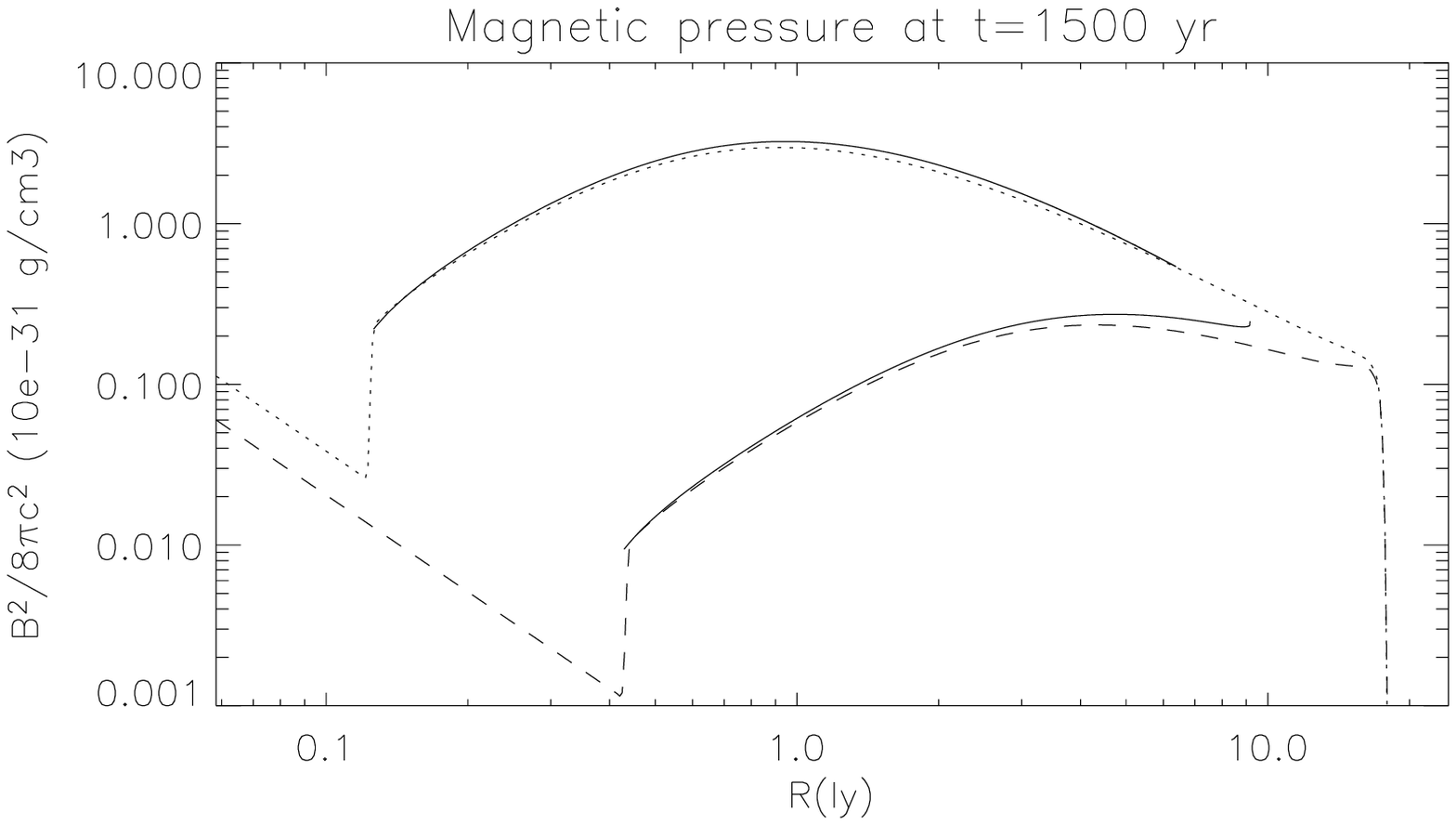}}
\caption{Comparison of the results with EC (solid line)for the $n=3$ case at t=1500 yr. Density, thermal pressure and magnetic pressure are shown for all the magnetizations we have considered: $\sigma=0.003$ (dotted line), $\sigma=0.0016$ (dashed line), $\sigma=0$ (dash-dotted line). The total pressure at the border is the same in the various cases. Comments are the same as for Fig.~\ref{fig:profn0}. }
\label{fig:profn3}
\end{figure}

In Figs.~\ref{fig:profn2} and \ref{fig:profn3} we plot the results of our simulations in the case $n=2$ and $n=3$ for all the different values of $\sigma$ we adopted. A comparison is made between the profiles that result from our simulations and those computed based on the EC model for the same wind magnetization. The value of $V_{ts}$ needed to compute the appropriate EC solution is derived again from Eq.~\ref{eq:presc}. Actually, our simulations give a generally lower value of $V_{ts}$. Moreover this decreases as the ram pressure drops, eventually becoming negative (the shock starts collapsing back to the pulsar in the case with $\sigma=0.003$, $n=3$ in Fig.~\ref{fig:evoln3}). However, the comparisons in Figs.~\ref{fig:profn2} and \ref{fig:profn3} are not improved by using the exact values of $V_{ts}$ given by the simulations.

Again we find that the self-similar model gives a reasonably good approximation of the simulation results in the post shock region up to the radius at which the magnetic pressure starts dominating over the thermal pressure, but fails, as could be expected, in the outer layers of the nebula. Moreover the singularity of the self-similar model, identified as the outer radius of the nebula to use for comparison with EC, is well inside the external boundary as determined from the simulations. The outer part of the nebula shows also a positive velocity gradient. This is the effect of the extra energy in the outer layer: this layer was created when the pulsar was more energetic and carries more energy than it would if the pulsar luminosity had stayed constant, so that it tends to expand and causes the material more recently injected by the wind to be confined at smaller radii. The radius of the EC singularity corresponds, within a 10~\% uncertainty, to the point where the speed of the flow is at its minimum.

\begin{figure}
\resizebox{\hsize}{!}{\includegraphics{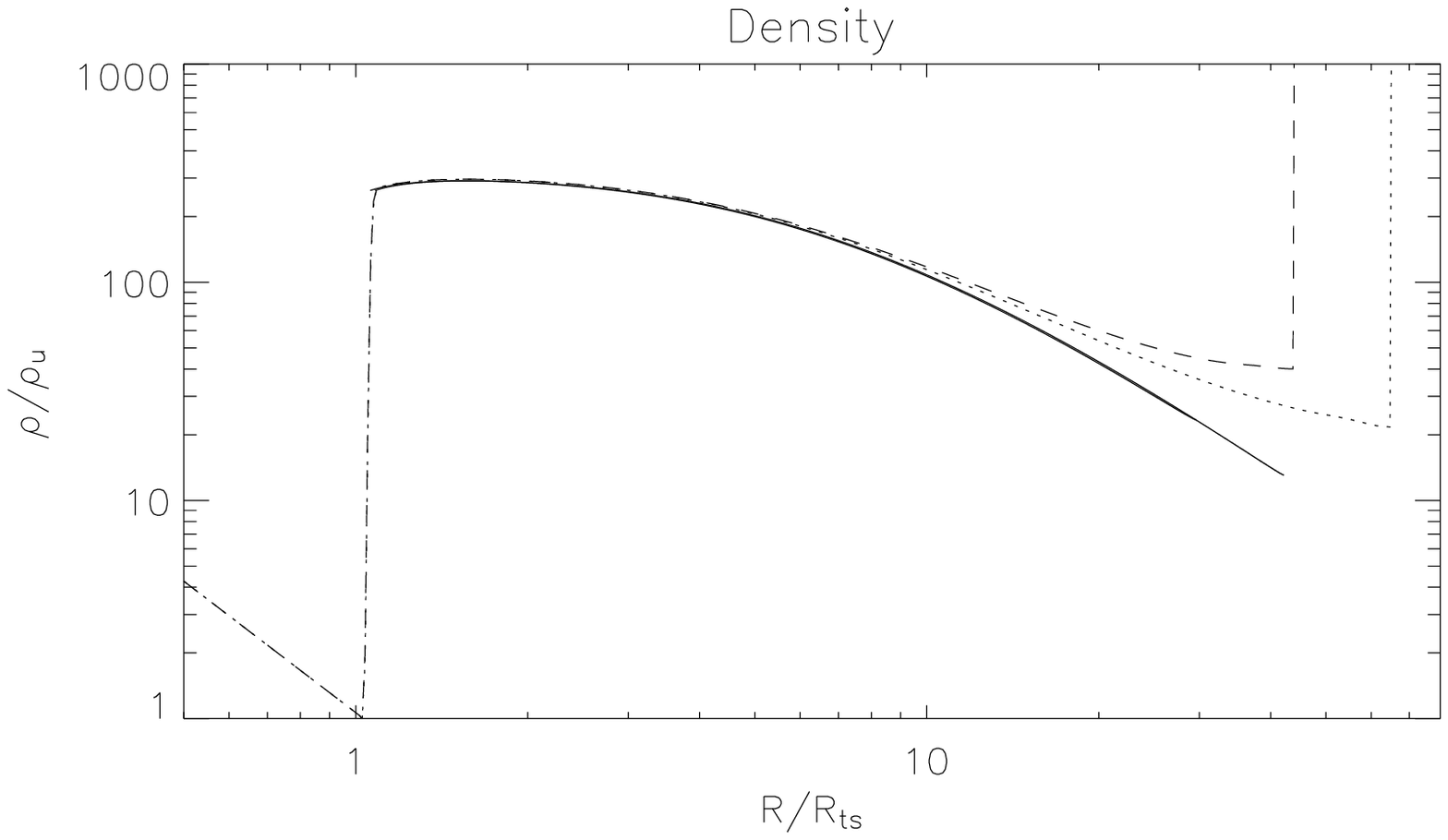}}
\resizebox{\hsize}{!}{\includegraphics{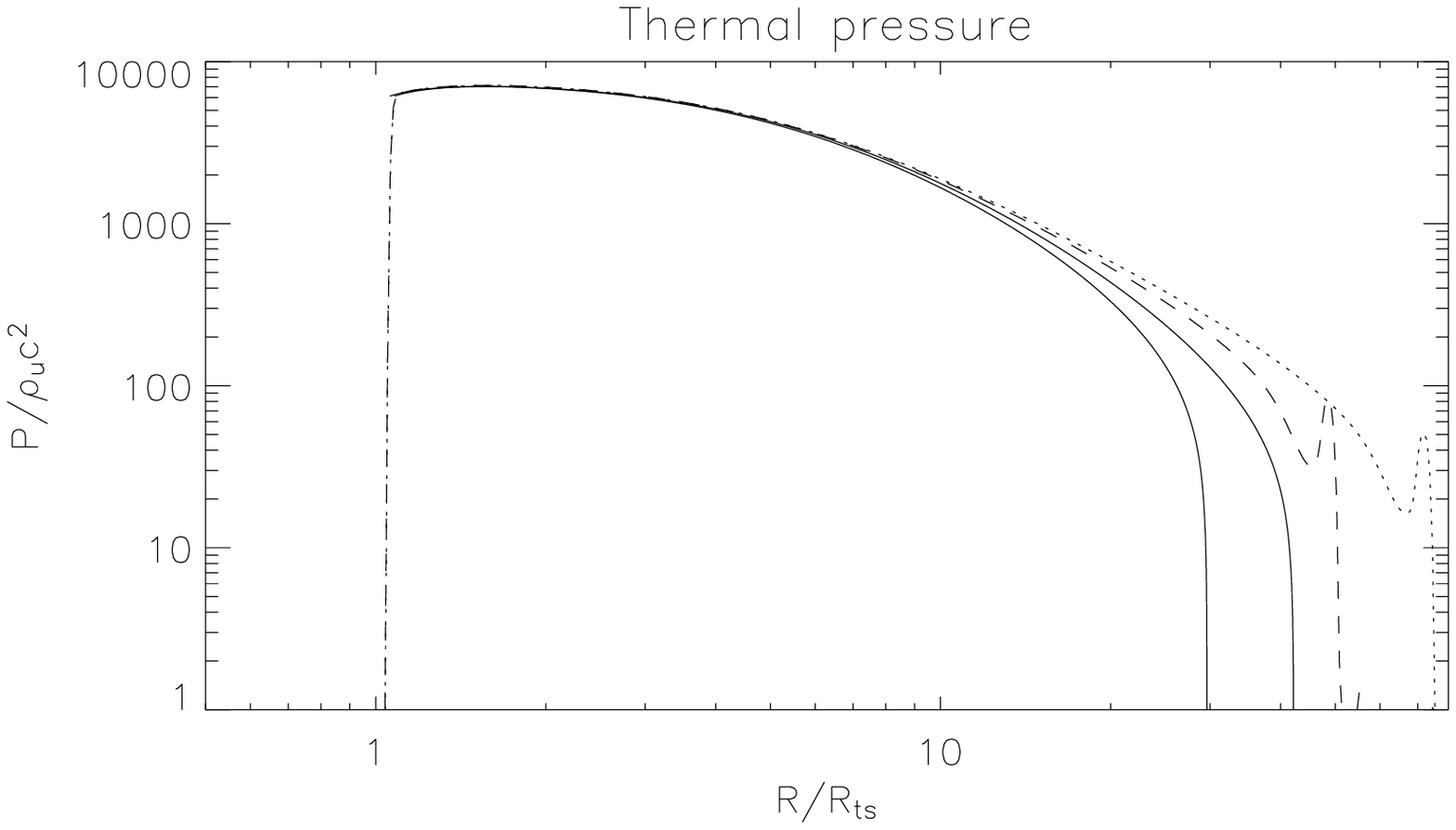}}
\resizebox{\hsize}{!}{\includegraphics{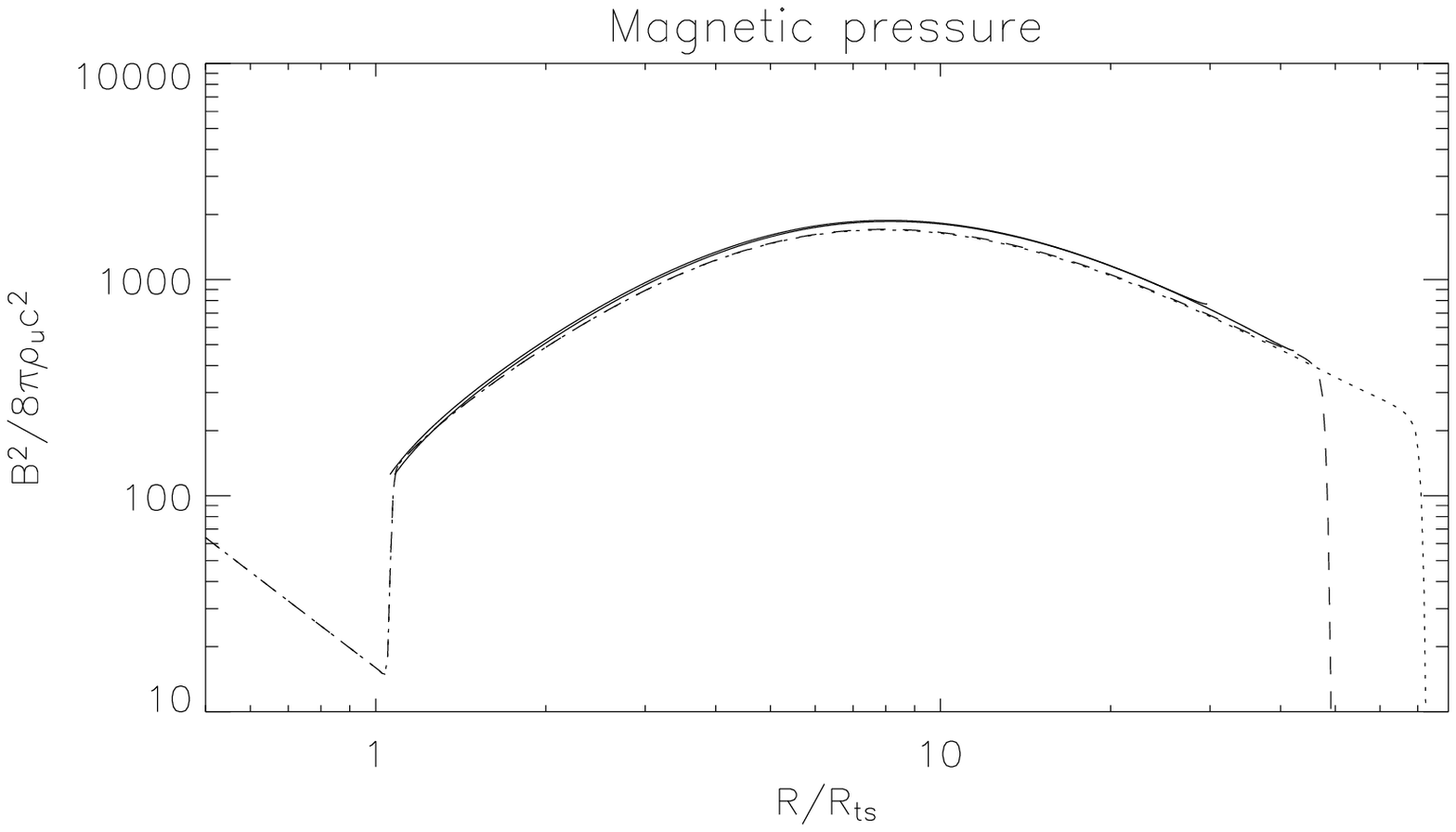}}
\caption{Comparison of the results in the $n=2$ case at t=1500 yr (dashed line) and in the case  $n=3$ case at t=760 yr (dotted line). The quantity $\rho_u$ that appears in the label of the $y$-axis is the density upstream of the termination shock. Radial distances are normalized to the termination shock radius. In the different panels we show, from top to bottom, the density, thermal pressure and magnetic pressure profiles. In all cases $\sigma=0.003$. The solid line represents the EC model. $\gamma =100$.}
\label{fig:compn2n3}
\end{figure}
In Fig.~\ref{fig:compn2n3} we compare the results obtained for $\sigma=0.003$ and $n=2$ and $n=3$ at the time when the pulsar luminosity is the same. Despite the different conditions, the various profiles coincide almost completely before the maximum of the magnetic pressure, and deviations appear only in the outer layer of the nebula. This suggests that the internal structure of a PWN is much more affected by the instantaneous properties of the wind, rather than by the overall pulsar history. Therefore, the appearance of the inner region of the PWN could in principle be used to estimate the value of $\sigma$ from non-thermal emission, without sensitivity to the spin-down process. Once the magnetization of the wind is known, the ratio $R_{cd}/R_{ts}$ could be used to infer the effect of spin-down (and eventually to determine $n$ and $\tau$).

\subsection{Spin-down factorization}
\label{sec:numres3}
\begin{figure}[h!!!!!!]
\resizebox{8cm}{!}{\includegraphics*{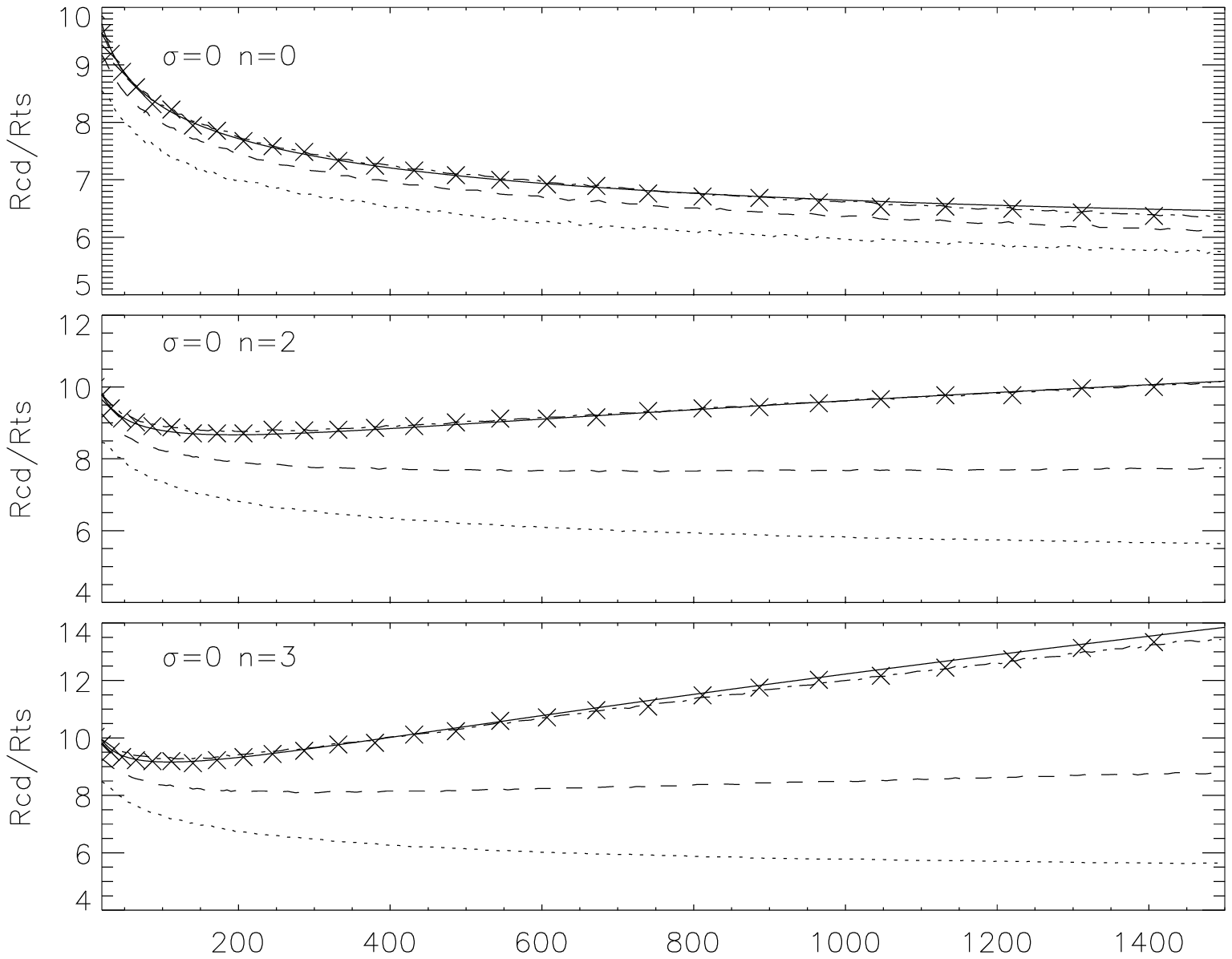}}\\
\resizebox{8cm}{!}{\includegraphics*{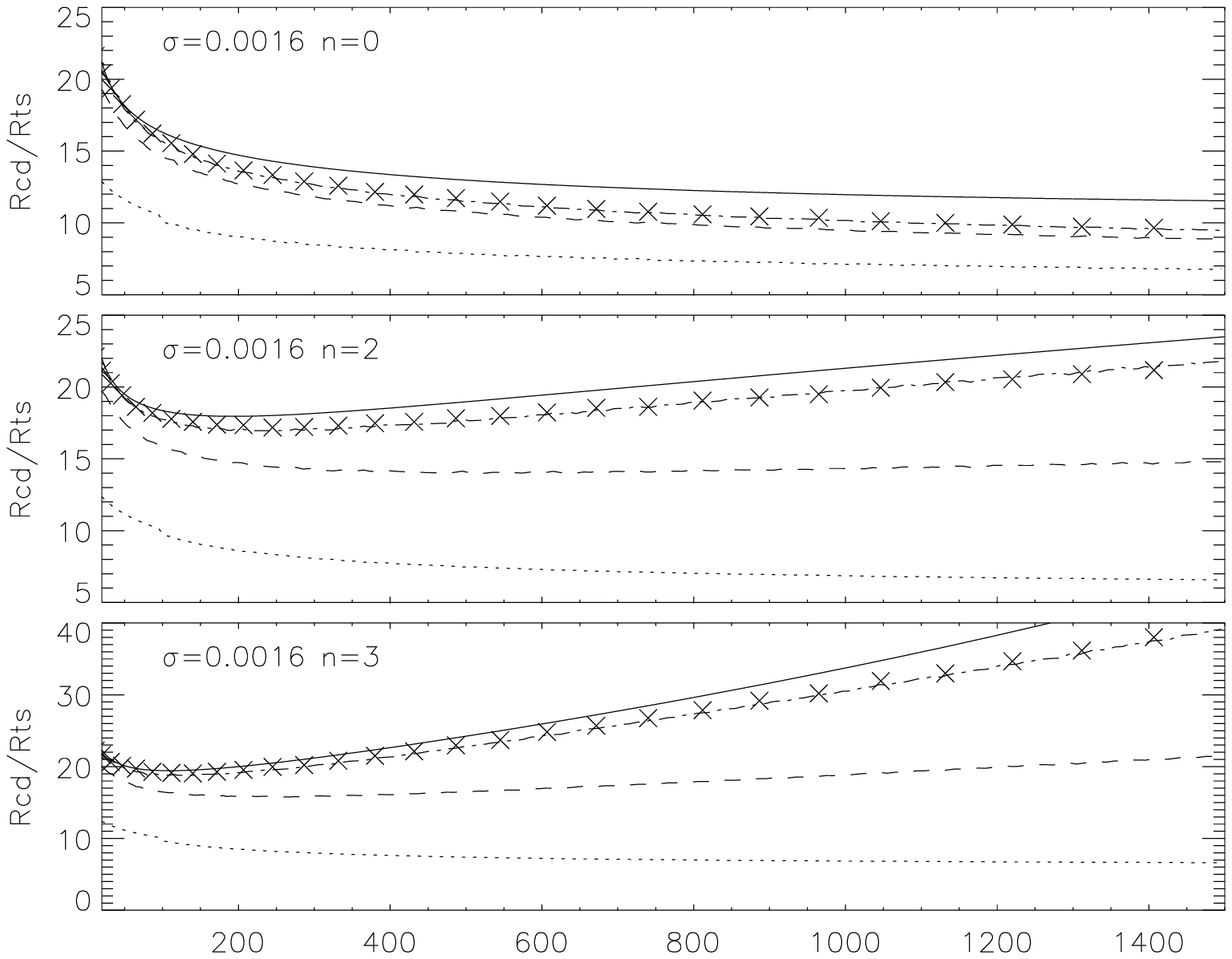}}\\
\resizebox{8cm}{!}{\includegraphics*{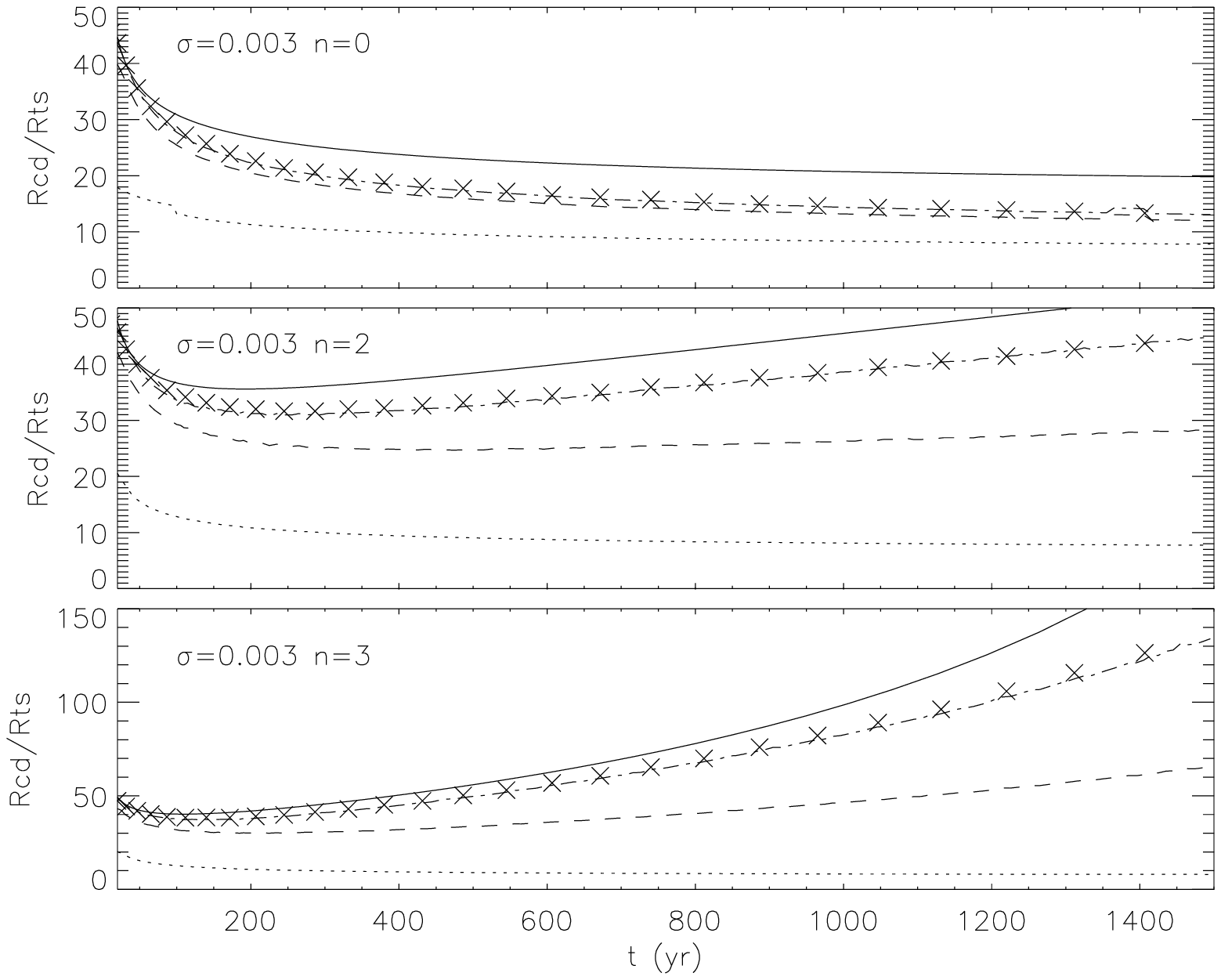}}
\caption{Evolution of the PWN size, in units of the termination shock radius. The upper three panels refer to the HD case, the middle three to $\sigma =0.0016$, and the lower three to $\sigma=0.003$. In each panel, the points refer to the result of our simulation; the dotted line is obtained using the value given by EC for the radius at which $v=V_{cd}$; the dashed line is the total size of the nebula in the EC model; the dot-dashed line is the fit done according to the correction in Eq.~\ref{eq:fit}; the solid line, finally, represents the solution of Eq.~\ref{eq:yt} normalized to the numerical results at time $t=50$ yr.}
\label{fig:size}
\end{figure}

In Fig.~\ref{fig:size} the evolution of the ratio $R_{cd}/R_{ts}$ is shown for the various values of $\sigma$ and $n$ employed. Neglecting the spin-down effect may lead to considerable errors in the description of the PWN, even for ages comparable or less than the characteristic spin-down time.

We want to stress that the ratio $R_{cd}/R_{ts}$ plays an important role, being commonly used to infer the wind magnetization and from this the PWN properties. If the effects of the spin-down are not taken into account, the magnetization of the wind is overestimated, and the profiles derived for the dynamically important quantities may be wrong.

If the value of the magnetization parameter $\sigma$ is not known from independent constraints than the radius of the termination shock, to apply consistently the analytic model described in Section~\ref{sec:theor} the values of $R_{cd}$ and $R_{ts}$ must be known at two different times, in order to fix both the normalization and $\sigma$ (see end of Section~\ref{sec:theor}). However in Subsection~\ref{sec:numres2} we have verified that the EC model provides a good description for $R_{cd}/R_{ts}$ in the case $n=0$ or equivalently $t \ll \tau $. One can then use the result of the EC model to make up for the lack of observations over long time lapses. Thus the problem of evaluating $\sigma$ and the termination shock evolution could be reduced to an eigenvalue problem to be solved iteratively.

From our simulations (see Fig.~\ref{fig:size}) we find that the best fit for the ratio $R_{cd}/R_{ts}$ is obtained multiplying the value given by EC (with $V_{ts}$ given in Eq.~\ref{eq:presc}) by
\be
C_{o}\left(\frac{(1+t/\tau)^{n}-(1+t/\tau)}{(n-1)t/\tau}\right)^{0.17+33\; \sigma}\ .
\label{eq:fit}
\ee
$C_{o}$ compensates for small differences that are present even in the case $n=0$, and its values are: $C_{o}=1.04$ in the HD case, and 1.06 and 1.07 for $\sigma=0.0016$ and $\sigma=0.003$ respectively. Notice that the effects of the magnetization are summarized in the exponent. It is possible in principle to use this equation together with Eq.~\ref{eq:presc} to estimate either $\sigma$ if $n$ and $\tau$ are known, or the spin-down parameters if the magnetization is independently known (for example from the comparison of synchrotron and Inverse Compton emission).

Neglecting adiabatic losses, the formula obtained for a generic $n$ is in shape similar to Eq.~\ref{eq:fit}, but with an exponent 0.5. Moreover Eq.~\ref{eq:fit} will be valid only for small values of $\sigma$ given the fact that the exponent cannot exceed the value 0.5 (no adiabatic losses). More generally as $\sigma$ increases the value of the exponent will tend to 0.5.

\subsection{The Crab Nebula}
\label{sec:numres4}
 As a special case we shall consider the Crab Nebula. This is surely the best studied PWN: a wealth of information is available but still no answer has been given to a number of problems.

 We want to point out that 1D models are an oversimplification in general. More so for the Crab Nebula, where very clear axisymmetry is observed in the X-rays (the renowned jet-torus structure). This has suggested that important deviations from the spherically symmetric approximation might occur in the pulsar wind region, leading to the formation of a turbulent flow in the nebula (e.g. \cite{komissarov03} 2003). 

In modeling the Crab Nebula, one important parameter is the actual value of the contact discontinuity velocity; recent estimates (\cite{fesen97}; \cite{sankrit97}; \cite{sollerma00}) give the value 1500 km/s with an upper limit of 2000 km/s. If we use the EC model without spin-down with the prescription in Eq.~\ref{eq:presc} we find that, in order to end up at the present time with a nebula that has $R_{cd}/R_{ts}=20$, a wind with $\sigma=0.0009$ or $\sigma=0.0015$ is required, depending on whether $V_{cd}=1500 {\rm km/s}$ or $V_{cd}=2000 {\rm km/s}$ is used. To include the effects of the spin-down, we use the following values: $\tau=720$ yr, $n=2.318$ (\cite{camilo00}; \cite{lyne93}; \cite{lyne88}). Using these values and the correction form given in Eq.~\ref{eq:fit}, we find for the magnetization of the Crab wind the value $\sigma=0.0004$, or, if the upper value of the expansion speed is used, $\sigma=0.0009$. 

We emphasize that these should be taken as estimates of the magnetization of the equatorial part of the outflow from the Crab pulsar, rather than being interpreted as reliable estimates of the overall magnetization of the Crab pulsar wind.

\section{Conclusions}
\label{sec:fin}
Our results show that, in the spherically symmetric approximation, the EC model gives a good description of the internal profiles of PWNe when the effects of spin-down can be neglected (i.e. when the central object is young and not very powerful). Noticeable discrepancies arise only at the outer boundary of the nebula, where the initial conditions and the evolutionary history of the system both play a major role. However, the region just behind the termination shock and further out, up to the position of the equipartition point, is well reproduced. This is the region of the nebula from which the high energy emission originates. 

Even if the radial profiles of the various quantities are in good agreement, the ratio $R_{cd}/R_{ts}$ inferred from the EC model by matching the boundary velocity for a given $\sigma$ is generally underestimated, due to the discrepancies at the boundary. Hence, if the ratio $R_{cd}/R_{ts}$ estimated in this way is compared with the observed value to evaluate the wind magnetization, one generally overestimates the value of $\sigma$. A better agreement is obtained if $R_{cd}$ is taken equal to the radius at which the singularity of the EC model is found.

There is a good agreement in the HD case while a small difference, less than $10\%$, is found in the magnetized cases, probably because in these cases $V_{cd} \sim V_{asy}$. The EC model can be adapted to the evolution of $V_{cd}$ since, with respect to the sound speed crossing time in the PWN, variations at the border are very slow: changes of $V_{cd}$ are compensated by analogous variations of $V_{ts}$.

More generally, we expect that deviations from the spherically symmetric approximation might occur in the pulsar wind region, leading to the formation of a turbulent flow in the nebula that only multidimensional simulations can handle. In spite of being a simplification of the problem, 1D models can help understanding the importance of various processes like spin-down, mass loading or synchrotron losses, and to clarify how to keep them into account when comparing models with observations.

Once the spin-down is included, the ratio $R_{cd}/R_{ts}$ increases with respect to the constant luminosity case. Again, if this effect is not taken into account, the pulsar wind magnetization inferred from the PWN size might well be overestimated. With a simplified analytic model we are able to reasonably reproduce the ratio $R_{cd}/R_{ts}$ given by the simulations. We also found a fit to the results of our simulations in the form of a correction coefficient that should multiply the standard EC expectation. However, the radial profiles of the various quantities given by the EC model fail to reproduce the structure inside the PWN beyond the point where magnetic pressure reaches its maximum.

Given our fit, if the wind magnetization is known from the analysis of the non-thermal emission of the PWN, the size of the nebula can be used, in principle, to estimate $n$ and $\tau$. Viceversa, if the pulsar spin-down properties are known, from the same expression it is possible to estimate $\sigma$. 

As shown in the case of the Crab Nebula, inclusion of the spin-down effect reduces the estimated $\sigma$ by about one order of magnitude with respect to the value found by KC. We want to stress that this is not just a dynamical problem related to the size of the nebula alone. A less magnetized wind creates a less magnetized nebula. Synchrotron losses are less efficient and high energy particles have a longer life and can move farther from the termination shock. This suggests that the interpretation of synchrotron maps should take into account the new results we have presented.

\begin{acknowledgements}
This work has been partly supported by the Italian Ministry for University and Research (MIUR) under grants Cofin2001 and Cofin2002, and partly by a SciDAC grant from the US Department of Energy High Energy and Nuclear Physics Program.
\end{acknowledgements}


\end{document}